\documentclass[aps,prd,twocolumn,amssymb,eqsecnum,showpacs,showkeyes,secnumarabic,graphics,floatfix,nofootinbib,tightenlines,longbibliography,superscriptaddress]{revtex4-1}

\usepackage{graphicx}
\usepackage{epsfig} 
\usepackage{bm}
\usepackage{cancel}  
\usepackage{dcolumn}  
\usepackage{amsmath} 
\usepackage{hhline} 
\usepackage[dvipsnames]{xcolor}
\usepackage[utf8]{inputenc}
\usepackage{cancel}
\usepackage[dvipsnames]{xcolor}  
\usepackage{float}
\usepackage{rotating}
\usepackage{longtable}
\usepackage[breaklinks=true,colorlinks=true,
linkcolor=blue,urlcolor=blue,citecolor=MidnightBlue,% PDF VIEW
bookmarks=true,bookmarksopenlevel=2]{hyperref}
\usepackage{mathrsfs} 

\def\apj{{Astroph.\@ J.\ }}

\def\aj{{Astron.\@ J.\ }}

\def\nat{{Nature\ }}

\def \beq  {\begin{equation}}
\def \eeq  {\end{equation}}
\def \ber  {\begin{eqnarray}}
\def \eer  {\end{eqnarray}}
\begin{document}
\newcommand{\newc}{\newcommand}

\newc{\be}{\begin{equation}}
\newc{\ee}{\end{equation}}
\newc{\ba}{\begin{eqnarray}} 
\newc{\ea}{\end{eqnarray}}
\newc{\bea}{\begin{eqnarray*}}
\newc{\eea}{\end{eqnarray*}}
\newc{\D}{\partial}
\newc{\ie}{{\it i.e.} }
\newc{\eg}{{\it e.g.} }
\newc{\etc}{{\it etc.} }
\newc{\etal}{{\it et al.}}
\newc{\lcdm}{$\Lambda$CDM }
\newc{\lcdmnospace}{$\Lambda$CDM}
\newcommand{\nn}{\nonumber}
\newc{\ra}{\Rightarrow}
\newc{\omm}{$\Omega_{m}$ }
\newc{\ommnospace}{$\Omega_{m}$}
\newc{\fs}{$f\sigma_8$ }
\newc{\fsz}{$f\sigma_8(z)$ }
\newc{\fsnospace}{$f\sigma_8(z)$}
\newc{\plcdm}{Planck/$\Lambda$CDM }
\newc{\plcdmnospace}{Planck15/$\Lambda$CDM}
\newc{\wlcdm}{WMAP7/$\Lambda$CDM }
\newc{\wlcdmnospace}{WMAP7/$\Lambda$CDM}
\newcommand{\fss}{{\rm{\it f\sigma}}_8}
\newcommand{\LP}[1]{\textcolor{red}{[{\bf LP}: #1]}}

\title{Hubble tension or a transition of the Cepheid SnIa calibrator  parameters?}
\author{Leandros Perivolaropoulos}\email{leandros@uoi.gr}
\affiliation{Department of Physics, University of Ioannina, GR-45110, Ioannina, Greece}
\author{Foteini Skara}\email{f.skara@uoi.gr}
\affiliation{Department of Physics, University of Ioannina, GR-45110, Ioannina, Greece}

\date {\today}  

\begin{abstract}
We re-analyze the Cepheid data used to infer the value of the Hubble constant $H_0$ by calibrating Type Ia supernovae. We do not enforce a universal value of the empirical Cepheid calibration parameters $R_W$ (Cepheid  Wesenheit color-luminosity parameter) and $M_H^{W}$ (Cepheid  Wesenheit H-band absolute magnitude). Instead we allow for variation of either of these parameters for each individual galaxy. We also consider the case where these parameters have two universal values: one for low galactic distances $D<D_c$ and one for high galactic distances $D>D_c$ where $D_c$ is a critical transition distance. We find hints for a $3\sigma$ level mismatch between the low and high galactic distance parameter values. We then use model selection criteria (AIC and BIC) which penalize models with large numbers of parameters, to compare and rank the following types of $R_W$ and $M_H^{W}$ parameter variations: Base models: Universal values for  $R_W$ and $M_H^{W}$ (no parameter variation), I: Individual fitted galactic $R_W$ with one universal fitted $M_H^{W}$, II: One universal fixed $R_W$ with individual fitted galactic $M_H^{W}$, III: One universal fitted $R_W$ with individual fitted galactic $M_H^{W}$,  IV: Two universal fitted $R_W$ (near and far) with one universal fitted $M_H^{W}$, V: One universal fitted $R_W$  with two universal fitted $M_H^{W}$ (near and far),  VI: Two universal fitted $R_W$ (near and far)  with two universal fitted $M_H^{W}$ (near and far). We find that the AIC and BIC  model selection criteria consistently favor model IV  instead of the commonly used Base model where no variation is allowed for the Cepheid empirical parameters. The best fit value of the SnIa absolute magnitude $M_B$ and of $H_0$ implied by the favored model IV is consistent with the inverse distance ladder calibration based on the CMB sound horizon $H_0=67.4\pm 0.5\,km\,s^{-1}\,Mpc^{-1}$. Thus in the context of the favored model IV the Hubble crisis is not present. This model may imply the presence of a fundamental physics transition taking place at a time more recent than $100\,Myrs$ ago.
\end{abstract}
\maketitle

\section{INTRODUCTION}
\label{INTRODUCTION}

A number of challenges of the standard $\Lambda$CDM model has been emerging during the past few years as the accuracy of cosmological observations improves. There are signals in cosmological and astrophysical data that appear to be in some tension ($2\sigma$ or larger) with the standard $\Lambda$CDM model as defined by the Planck18 parameter values \cite{Aghanim:2018eyx} (for a  review, see Ref. \cite{Perivolaropoulos:2021jda} and for a list of papers, see \cite{CosmologyGroupUOI}).

The most intriguing large scale tension is the Hubble crisis. Using a distance ladder approach, the local (late or low redshift)  measurements of the Hubble constant $H_0$ lead to values that are significantly higher than those inferred using the angular scale of fluctuations of the cosmic microwave background (CMB) in the context of the $\Lambda$CDM model. Local direct measurements of $H_0$ are in more than $4\sigma$ tension with CMB indirect measurements of $H_0$ (for a review, see Refs. \cite{Perivolaropoulos:2021jda,Verde:2019ivm,DiValentino:2021izs}). 

The local (late or low redshift)  determination of the Hubble constant $H_0$ using a distance ladder approach  depends on a chain of distance measurements. In the cosmic distance ladder approach each step of the distance ladder uses parallax methods and/or the known intrinsic luminosity of a standard candle source to determine the absolute (intrinsic) luminosity of a more luminous standard candle residing in the same galaxy. Thus highly luminous standard candles are calibrated for the next step in order to reach out to high redshift luminosity distances. If one of these distance measures is subject to systematics or new physics all the subsequent rungs of the cosmic distance ladder are off. 

The distance ladder approach is based on a method pioneered by Henrietta Swan Leavitt. She realized that a type of pulsating stars known as Cepheid variable has a period of pulsation that depends on its luminosity. This period–luminosity (PL) relation  is called  the Leavitt law \cite{Leavitt:1908vb,Leavitt:1912zz}. Knowing the luminosity of a Cepheid means that its luminosity distance can be determined just by observing its brightness which has been dimmed by that distance. Therefore the Cepheids whose luminosities are correlated  with their periods of variability can be the first standard candles in  the cosmic distance ladder \cite{Madore:1991yf,1992PASP..104..599J,1999PASP..111..775F,10.1086/341698,doi:10.1146/annurev.astro.43.072103.150612,Fouque:2003tm,Fouque:2007mx,Freedman:2010xv}. Trigonometric parallax methods (geometric anchors) may be used to calibrate the Cepheid variable star standard candles at the local Universe (primary distance indicators). Then using the measured luminosity distances of the calibrated Cepheid stars, the intrinsic luminosity of nearby ($D\approx 20-40\,Mpc$) incredibly bright type Ia supernova (SnIa) residing in the same galaxies as the Cepheids is obtained. This calibration of the new type of standard candle SnIa  fixes its absolute magnitude $M_B$ and is then used for SnIa at more distant galaxies (in the Hubble flow) to measure $H_0$ ($z\in [0.01,0.1]$) and $H(z)$ ($z\in [0.01,2.3]$) via the measurement of their luminosity distances.

The angular diameter distance of standard rulers can also be used for the estimation of $H(z)$. The $\Lambda$CDM best fit value obtained from the sound horizon at recombination  standard ruler calibrated by the CMB anisotropy spectrum peaks, leads to the measurement $H_0=67.4\pm0.5$ $km$ $ s^{-1} Mpc^{-1}$ \cite{Aghanim:2018eyx}. The local measurements based on Cepheid calibrators indicate the value $H_0=74.03\pm1.42$ $km$  $ s^{-1} Mpc^{-1}$ \cite{Riess:2019cxk} ($4.4\sigma$, $9\%$).  A more recent analysis by the Supernovae $H_0$ for the Equation of State (SH0ES) Team \cite{Riess:2011yx,Riess:2016jrr,Riess:2019cxk}  using the recently released Gaia Early Data Release $3$ (EDR$3$) parallaxes \cite{2021A&A...649A...1G} has reached  $1.8\,\%$ precision by improving the calibration. Thus, a value of $H_0= 73.2\pm 1.3$  was obtained \cite{Riess:2020fzl}. This is in a $4.2\sigma$ tension with the prediction from Planck18 CMB  observations \cite{Aghanim:2018eyx}. 

\begin{figure*}
\begin{centering}
\includegraphics[width=0.88\textwidth]{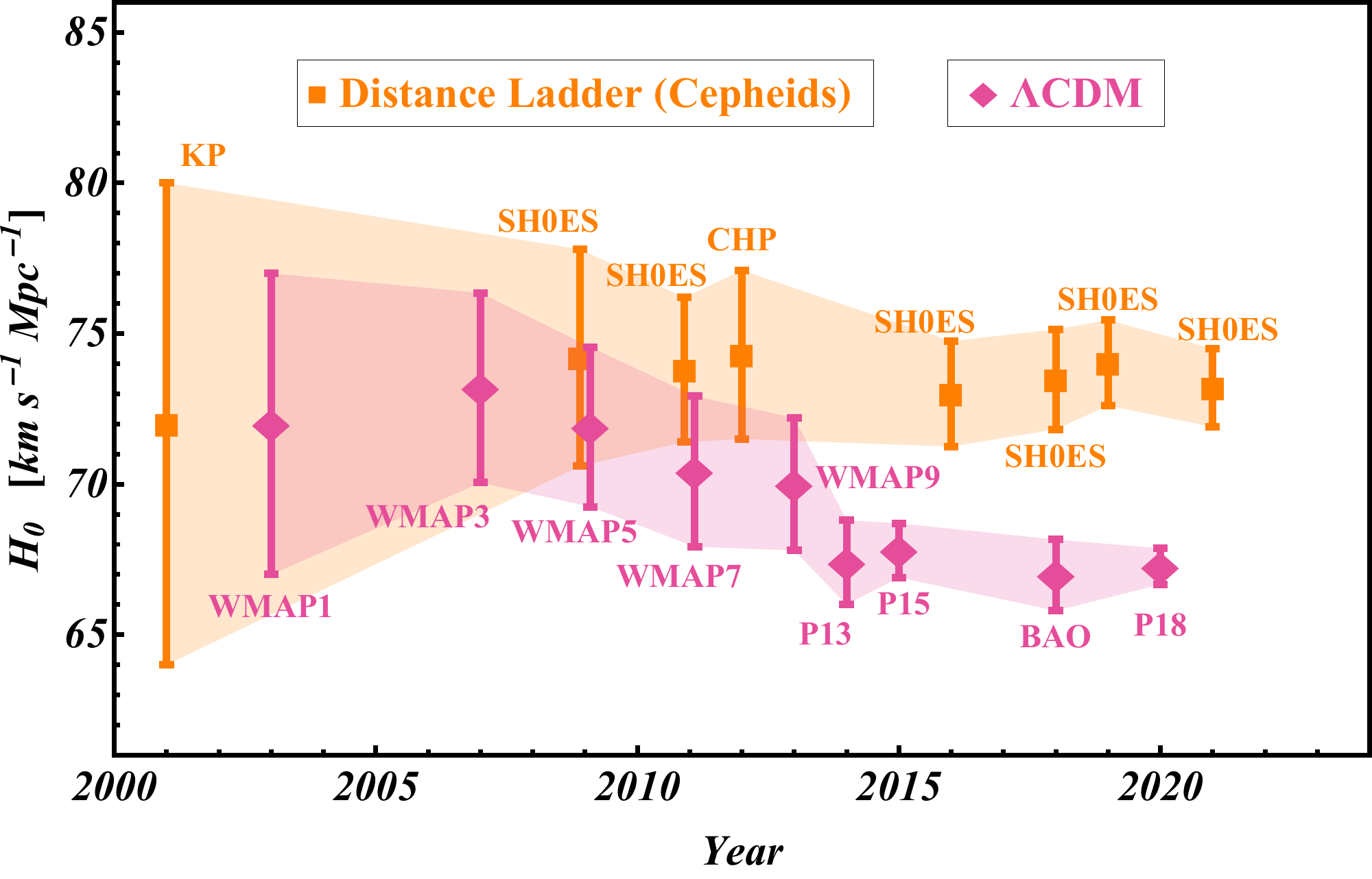}
\par\end{centering}
\caption{The Hubble constant as a function of publication date, using a set of different tools.  Symbols in orange denote the values of $H_0$ determined in the late Universe with a calibration based on the Cepheid distance scale (Key Project (KP) \cite{Freedman:2000cf}, SH0ES \cite{Riess:2009pu,Riess:2011yx,Riess:2016jrr,Riess:2018uxu,Riess:2019cxk,Riess:2020fzl}, Carnegie Hubble Program (CHP) \cite{Freedman:2012ny}). Symbols in purple denote the derived values of $H_0$  from analysis of the CMB data based on the sound horizon standard ruler (First Year WMAP (WMAP1) \cite{Spergel:2003cb}, Three Year WMAP (WMAP3) \cite{Spergel:2006hy}, Five Year WMAP (WMAP5) \cite{Dunkley:2008ie}, Seven Year WMAP (WMAP7) \cite{Komatsu:2010fb}, Nine Year WMAP (WMAP9) \cite{Bennett:2012zja}, Planck13 (P13) \cite{Ade:2013zuv}, Planck15 (P15) \cite{Ade:2015xua}, Planck18 (P18) \cite{Aghanim:2018eyx}, BAO \cite{Addison:2017fdm}). The orange and purple shaded regions demonstrate the evolution of the uncertainties in these values which have been decreasing for both methods. The most recent measurements disagree at greater than $4\sigma$ (from Ref. \cite{Perivolaropoulos:2021jda}).}
\label{figh0year}
\end{figure*}

\begin{figure*}
\begin{centering}
\includegraphics[width=0.92\textwidth]{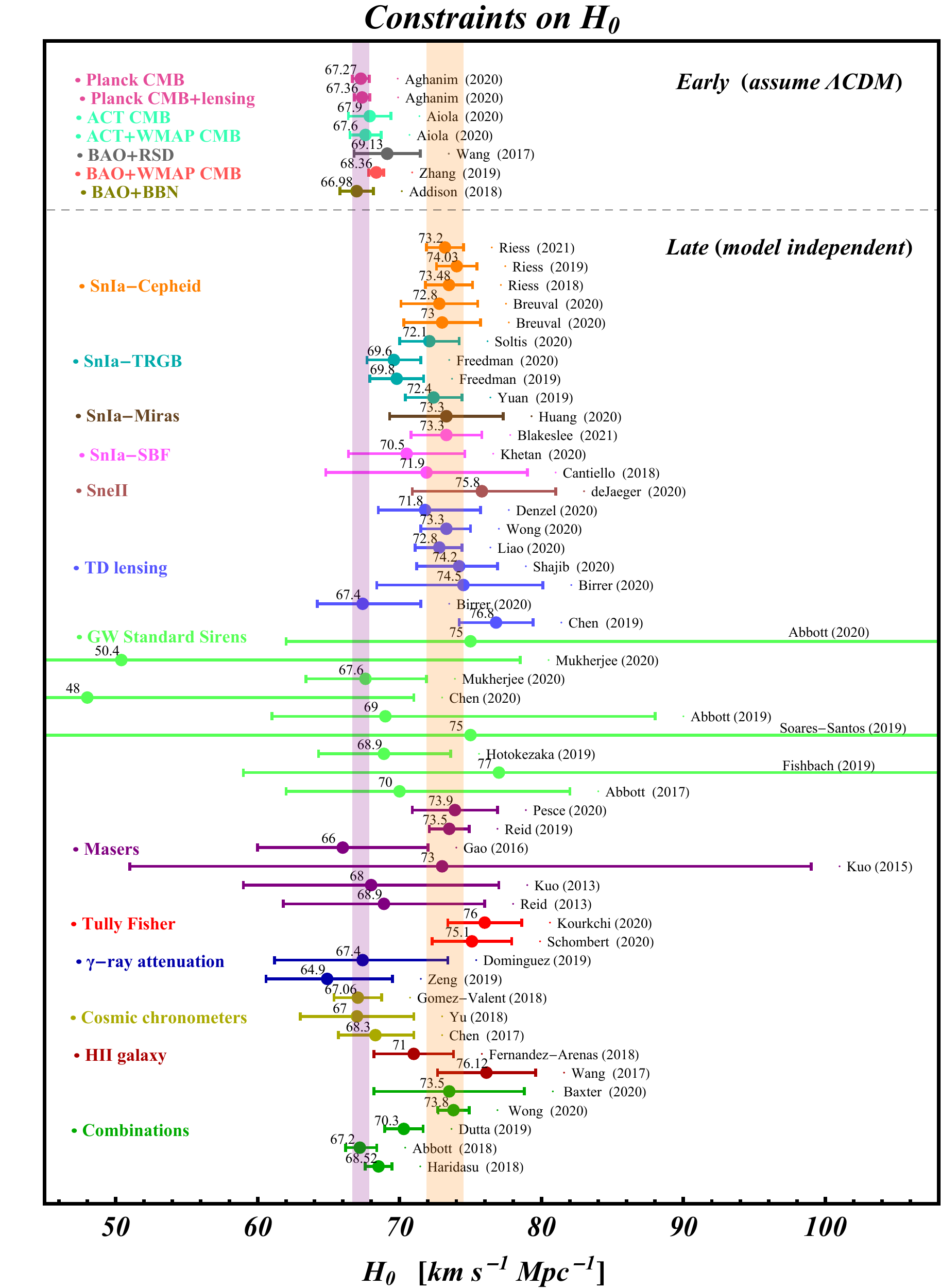}
\par\end{centering}
\caption{The Hubble constant $H_0$ values with the $68\%$ CL constraints derived by recent measurements. The value of the Hubble constant $H_0$ is derived by early time approaches based on sound horizon, under the assumption of a $\Lambda$CDM background. The shaded orange  and purple bands show the tension  between the values inferred from Cepheid calibrated SnIa and CMB observation (from Ref. \cite{Perivolaropoulos:2021jda}).}
\label{figh0}
\end{figure*}

The values of $H_0$ determined in the late Universe with a calibration based on the Cepheid distance scale and the derived values of $H_0$ from analysis of the CMB anisotropy spectrum data are shown in Fig. \ref{figh0year}. The uncertainties in these values have been decreasing for both methods and the recent measurements disagree beyond $4\sigma$. The Hubble constant $H_0$ values at $68\%$ CL through direct and indirect measurements obtained by the different methods are illustrated in Fig. \ref{figh0}, from
which is evident that the SnIa distance scale  calibrated by Cepheid variables is in tension with the CMB sound horizon scale.
 
If the Hubble tension is not due to systematic errors \cite{Efstathiou:2013via,Cardona:2016ems,Zhang:2017aqn,Follin:2017ljs,Feeney:2017sgx,Wu:2017fpr,Verde:2019ivm}, it could be an indication of incorrect estimate of the sound horizon scale due e.g. to early dark energy \cite{Chudaykin:2020acu} or to late phantom dark energy \cite{Alestas:2020mvb}. A Gpc scale underdensity \cite{Haslbauer:2020xaa} has also been proposed as a possible cause of the Hubble tension. Theoretical models addressing this discrepancy  utilize either a recalibration of the $\Lambda$CDM standard ruler (the sound horizon) assuming new physics before the time of recombination \cite{Karwal:2016vyq,Poulin:2018cxd,Agrawal:2019lmo,Ye:2020btb} or a deformation of the Hubble expansion rate $H(z)$ at late times \cite{Alestas:2020mvb,DiValentino:2016hlg,Caldwell:2005xb} or an abrupt transition of the SnIa absolute luminosity due to late time new physics \cite{Marra:2021fvf} (for a review, see Refs. \cite{Perivolaropoulos:2021jda,Verde:2019ivm,DiValentino:2021izs}, for approaches, see Refs. \cite{Kazantzidis:2019dvk,Alestas:2021xes} and for a relevant talk, see Ref. \cite{pres1}).

The authors of Ref. \cite{Marra:2021fvf} have pointed out that the $H_0$ crisis may be viewed as a mismatch between the SnIa absolute magnitude calibrated by Cepheids at $z<0.01$  \cite{Camarena:2019moy,Camarena:2021jlr}
\be
M_B^< = -19.244 \pm 0.037 \,mag
\label{maglate}
\ee
 and the SnIa absolute magnitude  obtained using the parametric free inverse distance ladder calibrating SnIa absolute magnitude using the sound horizon scale \cite{Camarena:2019rmj}
 \be
 M_B^>=-19.401\pm 0.027 \, mag
 \label{magearly}
 \ee
Since the two measurements are obtained at different redshifts they may indicate a transition in the absolute magnitude with amplitude $\Delta M_B=M_B^>-M_B^<\simeq -0.2\, mag$  at a transition redshift $z_t\lesssim 0.01$ (roughly 100-150 million years ago)  \cite{Alestas:2020zol,Marra:2021fvf}. Note that hints of a possible weak evolution of the absolute magnitude $M_B$ have been recently pointed out in Refs. \cite{Kazantzidis:2020xta,Kazantzidis:2020tko}.

Such a transition may occur due to a transition of the strength of the gravitational interactions $G_{\rm eff}$ which modifies the SnIa intrinsic luminosity $L$ by changing the value of the Chandrasekhar mass $M_{Ch}$. The simplest assumption leads to \cite{Amendola:1999vu,Gaztanaga:2001fh}
\be 
L\sim M_{Ch}\sim G_{\rm eff}^{-3/2}
\label{lmg}
\ee
even though corrections may be required to the above simplistic approach \cite{Wright:2017rsu}. 

Using the normalized effective Newton constant $\mu_G\equiv G_{eff}/G_N$ (where $G_N$ is the locally measured Newton’s constant and $\mu_G=1$ for $z<z_t\lesssim 0.01$) and assuming the power law dependence Eq. (\ref{lmg}) the absolute magnitude of SnIa $M_B$ is expected to
change as \cite{Garcia-Berro:1999cwy,Gaztanaga:2001fh,Nesseris:2006jc,Kazantzidis:2018jtb}
\be
\Delta M_B=\frac{15}{4}\log_{10}(\Delta \mu_G -1)
\ee
Thus for $\Delta M_B\simeq -0.2\,mag$ it is straightforward to show that the change of $\mu_G$ \cite{Marra:2021fvf} 
\be
\Delta \mu_G \equiv\mu_G^>-\mu_G^<\simeq -0.12
\ee
where $\mu_G^>$ corresponds to $z>z_t$ and   $\mu_G^<$ corresponds to $z<z_t$ in the context of a $\Lambda$CDM background $H(z)$. 

This connection indicates that a $10\%$ smaller $G_{\rm eff}$ for $z>z_t$ could potentially provide the required decrease of the absolute magnitude (increase of luminosity) at early times for the resolution of the Hubble tension. In fact such a decrease would also lower the growth rate of cosmological perturbations thus helping in the resolution of the growth tension \cite{Marra:2021fvf,Heymans:2020gsg} . Thus, the smaller $G_{\rm eff}$ ($G_{eff}<1$) for $z>z_t$  hints towards weaker gravity as indicated by studies discussing the growth tension \cite{Nesseris:2017vor,Amendola:2017orw,Gannouji:2018ncm,Kazantzidis:2018rnb,Perivolaropoulos:2019vkb,Skara:2019usd,Gannouji:2020ylf}.

Interestingly, an analysis of recent Tully-Fisher data \cite{Alestas:2021nmi} has identified hints for a transition of the values of the Tully-Fisher parameters which depend on gravitational physics at a distance scale close to $20\,Mpc$. The magnitude and sign of the transition are consistent with the proposed mechanism albeit at somewhat more recent times: about 70 million years ago, corresponding to $z_t\simeq 0.005$. 

A recent analysis \cite{Mortsell:2021nzg}, has analyzed the color-luminosity relation of Cepheids in anchor galaxies and SnIa host galaxies by identifying the color-luminosity relation for each individual galaxy instead of enforcing a universal color-luminosity relation to correct the near infrared (NIR) Cepheid magnitudes.  This analysis finds a systematic brightening of Cepheids at distances larger than about $20\, Mpc$ (see Fig. 4 in Ref. \cite{Mortsell:2021nzg}). As pointed out in \cite{Mortsell:2021nzg} this brightening could be enough to resolve the Hubble tension. The authors attribute it to variation of dust properties but there is currently a debate on the actual cause of this brightening. 

In the present study we reproduce and extend the analysis of Ref. \cite{Mortsell:2021nzg} by considering a varying among individual galaxies color-luminosity relation and (in a separate analysis) allowing the Cepheid absolute magnitude $M_H^W$ to vary among individual galaxies instead of enforcing a universal absolute magnitude. We also explore the possibility that there are two universal absolute magnitudes: one applicable for low distance $D$ Cepheids ($M_H^{W,<}$ for $D<D_c$) and a second, applicable for high distance $D$ Cepheids ($M_H^{W,>}$ for $D>D_c$). We then test the consistency among the two absolute magnitudes searching for hints of a physics transition at some critical distance (time) $D_c$ ($D_c/c$).

We thus address the following questions:

\begin{itemize}
\item
Are there indications for variation of the color-luminosity relation and of the Cepheid absolute magnitude among individual galaxies?
\item
Is the color-luminosity relation and/or absolute magnitude of nearby Cepheids consistent with the corresponding properties of Cepheids in more distant galaxies?
\item
Are there indications for a Cepheid luminosity transition similar to the one required for the resolution of the Hubble tension?
\end{itemize}

In order to address these questions we use the same data and similar method as those used in Ref.  \cite{Mortsell:2021nzg} but in addition we extend the types of parameter variations allowed while implementing model selection criteria in order to compare the different allowed types of parameter variations (models) with the Base model which assumes universal Cepheid empirical parameter values. The data are obtained from Refs. \cite{Riess:2020fzl,Riess:2019cxk,Riess:2016jrr} and displayed in the Appendix \ref{DATA USED IN THE ANALYSIS}. Our generalized approach is based on two extensions
\begin{itemize}
    \item We break the assumption of universality not only on the  color-luminosity relation but also on the absolute magnitude of Cepheids.
    \item In addition to fitting the Cepheid color-luminosity relation (or the absolute magnitude) for each galaxy separately, we also consider the case of two universality classes one for nearby and one for more distant galaxies thus introducing only one new parameter in each case compared to the standard universal approach.
\end{itemize}

%Also earth paleontology data may lead to interesting constraints on the proposed mechanism.  A crude approach to this problem, discussed in \cite{Uzan:2010pm} and originally proposed in \cite{Teller:1948zz}, indicates that the temperature on Earth varies according to $G^{2.25}$, assuming that the mass of the Sun remains constant and ignoring atmospheric and other climatological effects (green-house effects etc). Within this approximation, if G had been smaller by $10\%$ $100–150$ million years ago, the temperature of the Earth would have been about $20\%$ lower.  Transitions in the temperature of the Earth of this magnitude are known to have taken place during the last $500$ million years, and are attributed to climatological effects or to the meteorite that is believed to have led to the extinction of various species, including the dinosaurs, about $70$ million years ago.  Therefore, if atmospheric effect uncertainties are taken into account, a $10\%$ gravitational transition cannot be easily a priori excluded.

%Our analysis is closely related to the results of study by Ref. \cite{Mortsell:2021nzg}. Using the same datasets we  search for a transition of the  best fit parameter values between cepheid data subsamples at low and high distances.  

The structure of this paper is the following: In the next Section \ref{METHOD-DATA} we present a brief review of the theoretical background and we describe the method used in our analysis as well as the data considered. In Section \ref{SEARCH FOR TRANSITION} we present our results focusing on the consistency of nearby and more distant samples with each other. We also compare with a Monte Carlo uniformized dataset in order to verify that any observed peculiar signal disappears in a Monte Carlo constructed homogeneous dataset. In Section \ref{selection} we consider various cases (models) allowing for different types of empirical parameter variation and use criteria for model selection and model comparison. In Section \ref{Transition as a possible solution of Hubble tension} we investigate the impact of the allowed types of parameter variation on the  SnIa absolute magnitude $M_B$ and on the corresponding derived value of $H_0$. Finally in Section \ref{CONCLUSION-DISCUSSION} we summarize our methods and results and discuss possible extensions of our analysis. We also compare our results with previous analyses searching for similar transition effects in different datasets.

\section{THEORETICAL BACKGROUND-METHOD-DATA}
\label{METHOD-DATA}

In this section, we present a brief review of the theoretical expressions  and describe the  method and the dataset used.

\subsection{Standard candles}  
\label{Standard candles} 

In an expanding flat Universe, where the energy is not conserved due to the increase of the photon wavelength and period with time, the luminosity distance can be expressed as \cite{Dodelson:2003ft,Perivolaropoulos:2006ce}
\be 
d_L(z)=c(1+z)\int_0^z \frac{dz'}{H(z')}
\label{dlz}
\ee
The luminosity distance is  an  important  cosmological observable that is measured using standardizable candles like SnIa ($z<2.3$) \cite{Scolnic:2017caz} and gamma-ray bursts (GRBs) ($0.1<z \lesssim 9$)  \cite{Schaefer:2006pa,Tanvir:2009zz,Cucchiara:2011pj,Wang:2015cya,Demianski:2016zxi,Tang:2019wlb}. 

Surveys can indicate the distance-redshift relation of SnIa by measuring their peak luminosity that is tightly correlated with the shape of their characteristic light curves (luminosity as a function of time after the explosion)  \cite{Phillips:1993ng} and the redshifts of host galaxies.

The use of SnIa in the measurement of $H_0$ and $H(z)$ relies on a basic assumption that they are standardizable and after proper calibration they have a fixed absolute magnitude independent of redshift in the redshift range $z\in [0.01,2.3]$ \cite{Phillips:1993ng}. This assumption has been tested in Refs. \cite{Colgain:2019pck,Kazantzidis:2019dvk,Kazantzidis:2020tko,Sapone:2020wwz,Koo:2020ssl,Kazantzidis:2020xta,Lukovic:2019ryg,Tutusaus:2018ulu,Tutusaus:2017ibk,Drell:1999dx}. The possibility for intrinsic luminosity evolution of SnIa with redshift was first highlighted by Ref. \cite{1968ApJ...151..547T}. Also, the assumption that the luminosity of SnIa is independent of host galaxy properties (e.g. host age, host morphology, host mass) and local star formation rate has been discussed in Refs. \cite{Kang:2019azh,Rose:2019ncv,Jones:2018vbn,Rigault:2018ffm,2018ApJ...854...24K}. 

The apparent magnitude $m_B$ of SnIa at redshift $z$ with absolute magnitude $M_B$ in the context of a specified form of $H(z)$, is related to their luminosity distance $d_L(z)$  in Mpc as
\be 
m_B(z)=M_B+5\log_{10}\left[\frac{d_L(z)}{Mpc}\right]+25
\label{apmagd}
\ee
Using now the dimensionless Hubble free luminosity distance
\be 
D_L(z)=\frac{H_0d_L(z)}{c}
\label{Dlz}
\ee
the apparent magnitude can be written as
\be 
m_B(z)=M_B+5\log_{10}\left[D_L(z)\right]+5\log_{10}\left[\frac{c/H_0}{Mpc}\right]+25
\label{apmag}
\ee
Using the degenerate combination 
\be 
\mathcal{M}=M_B+5\log_{10}\left[\frac{c/H_0}{Mpc}\right]+25
\label{combM}
\ee 
into Eq. (\ref{apmag}) we obtain
\be 
m_B(z)=\mathcal{M}+5\log_{10}\left[D_L(z)\right]
\label{aparmagz}
\ee
The use of Eq. (\ref{apmag}) to measure $H_0$ using the measured apparent magnitudes of SnIa requires knowledge of the value of the SnIa absolute magnitude $M_B$. This can be obtained using calibrators of local SnIa at $z<0.01$ (closer than the start of the Hubble flow) in the context of a distance ladder approach (e.g. \cite{Sandage:2006cv}). Calibrators like Cepheid stars which are bright, variable supergiants are used in this context.

\subsection{Cepheid calibration}
The Milky Way (MW),  the Large Magellanic Cloud (LMC) and NGC 4258 are used as distance anchor galaxies.
For Cepheids in the anchor galaxies there are three different ways of geometric distance calibration of their luminosities: parallaxes in the MW \cite{Benedict:2006cp,vanLeeuwen:2007xw,Casertano:2015dso,Riess:2014uga,2016A&A...595A...4L,Riess:2018uxu,Riess:2018byc,Riess:2020fzl}, detached eclipsing binary stars (DEBs) in the LMC \cite{2019Natur.567..200P} and water  masers\footnote{Very long baseline interferometric (VLBI) observations of water megamasers which are found in the accretion disks around supermassive black holes (SMBHs) in active galactic nuclei (AGN) have been demonstrated to be powerful one-step geometric probes for measuring extragalactic distances \cite{Herrnstein:1999kd,Humphreys:2013eja}.} in NGC 4258 \cite{Reid:2019tiq}. The DEBs method relies on surface-brightness relations and is one-step distance determination to nearby galaxies independent from Cepheids \cite{Pietrzynski:2013gia}. The Andromeda galaxy (M31) could serve as an anchor in the cosmic distance ladder but the uncertainty in its distance measurements is difficult to reduce \cite{Li:2021qkc}.

The empirically-determined period-magnitude relationship of a Cepheid can be expressed as (e.g. \cite{Follin:2017ljs})
\be 
m_H-R_HE(V-I)=\mu+M_H+b_H[P]+Z_H[M/H]
\label{permagrel}
\ee
where $m_H$ is the observed apparent magnitude in the near-infrared $H$ (F160W) band which is centered at $\lambda_H\sim 1.63\,\mu m$, $V$ (F555W) and $I$ (F814W) are the optical mean apparent magnitudes which are centered at $\lambda_V\sim 0.555\,\mu m$ and $\lambda_I\sim 0.79\,\mu m$ respectively, in the HST system\footnote{HST uses the same three-band photometric system with the Wide Field Camera 3  (WFC3) with two optical filters (F555W and F814W) and one near-infrared filter (F160W). }, $E(V-I)$ is the color excess, $R_H$ is the total to selective extinction  parameter at H band\footnote{The total to selective extinction parameter at H band $R_H$ is defined as $R_H\equiv\ A_H/(A_V-A_I)= A_H/E(V-I)$ (where $A_H$ is the extinction due to dust along the line of sight).}, $\mu \equiv 5\log_{10}\left[d_L(z)/Mpc\right]+25$ is the inferred distance modulus to the Cepheid, $M_H$ is the absolute magnitude of a period $P = 10\,d$  Cepheid ($d$ for days), $b_H$ and $Z_H$ are the slope parameters that represent  the dependence of magnitude on both period and metallicity. 

The $[M/H]$ is a measure of the metallicity of the Cepheid. The usual bracket shorthand notation for the metallicity $[M/H]$ represents the Cepheid metal abundance compared to that of the Sun 
\be
[M/H]\equiv \log(M/H)-\log(M/H)_{\odot}=\Delta \log(M/H)
\ee
where M and H is the number of metal (any element other than hydrogen or helium) and hydrogen atoms per unit of volume respectively. The unit often used for metallicity is the dex (decimal exponent) defined as $n\, dex = 10^n$. 

Also, the bracket shorthand notation for the period $[P]$ is used as  ($P$ in units of days)  
\be
[P]\equiv \log P-1    
\ee

The color excess characterizes the amount of reddening associated with interstellar extinction, a combined effect of absorption and scattering of blue more than red light by dust and other matter \cite{Draine:2003if,carroll_ostlie_2017}. The color excess depends on the properties of dust and is defined as
\be
E(V-I)\equiv A_V-A_I= (V-I)-(V-I)_0
\label{colorex}
\ee
where $V-I$ and $(V-I)_0$ are the observed and the intrinsic (normal) Cepheid color respectively. The latter is the hypothetical true Cepheid color which would be observed if there was no extinction.

Following the same formulation  used by the SH$0$ES team \cite{Riess:2016jrr,Riess:2019cxk} in order to minimize the impact of extinction correction uncertainties for distance measurements and  determination of the  Hubble constant $H_0$,  we use the replacement $E(V-I) \rightarrow V-I$ in Eq. (\ref{permagrel}), the Hubble Space Telescope (HST) NIR H-band and the reddening-free "Wesenheit" magnitudes \cite{1982ApJ...253..575M} 
\be 
m_H^W\equiv m_H-R_W(V-I)=\mu+M_H^W+b_W[P]+Z_W[M/H]
\label{wesmag}
\ee
where the empirical parameter $R_W$ is the reddening-free "Wesenheit" color ratio and is different from $R_H$ which  can be derived from a dust law (e.g. the Fitzpatrick law \cite{Fitzpatrick:1998pb}). The parameter $R_W$ corrects for both dust and intrinsic variations applied to observed blackbody colors $V-I$. Eq. (\ref{wesmag}) can be derived from Eq. (\ref{permagrel}) using Eq. (\ref{colorex}) with a constant fixed parameter $R_W$ under the important assumption that the intrinsic Cepheid color $(V-I)_0$ can be assumed to have the same distribution for all galaxies. This allows the absorption of the term $R_H (V-I)_0$ by the Cepheid absolute magnitude $M_H$ thus defining the Cepheid  Wesenheit H-band absolute magnitude $M_H^W$ in Eq. (\ref{wesmag}). An additional logarithmic dependence of the intrinsic color on the Cepheid period is also allowed and would be absorbed by the parameter $b_H$ leading to the parameter $b_W$. Thus, if there was a transition in the intrinsic Cepheid color $(V-I)_0$ at a given galactic distance, this transition would manifest itself as a shift of any or all of the parameters  $M_H^W$, $b_W$ and/or $R_W$ at the same distance. 

For distances based on NIR ($1< \lambda< 2.5\, \mu m$) measurements, both the impact of extinction by dust (gauged using the observed color $V-I$) and the impact of metallicity on the luminosities and colors of Cepheids, are less significant compared to the corresponding optical measurements \cite{1982ApJ...257L..33M,Madore:1991yf,1991ApJ...372..455F,Macri:2001it} (for the metallicity effects which are still largely debated, see Refs. \cite{2017ApJ...842..116W,2018A&A...620A..99G,2018A&A...619A...8G,2020A&A...642A.230R,2021ApJ...913...38B}). However the NIR measurements of Cepheids still suffer from crowding and blending (photometric contamination) from redder Red Giant Branch (RGB) and Asymptotic Giant Branch (AGB) disk stars, particularly as the distance increases \cite{Freedman_2019,Riess:2020xrj,Freedman:2021ahq}.

Wesenheit magnitudes have the advantage of smaller dispersion in the PL relation caused by differential extinction and the nonzero temperature width of the Cepheid instability strip\footnote{The instability strip refers to a narrow, almost vertical temperature (spectral type) region on the Hertzsprung-Russell (H-R) diagram which contains several types of pulsating variable stars including Cepheid variables (see e.g. Refs. \cite{1990ApJ...354..295F,Sandage:2008hg,2017ApJ...842...42M}). The Classical Cepheids (Population I Cepheids) are F6-K2 type supergiants ($\sim 4-20\,M_{\odot}$) with a period of 1 to 70 days with an amplitude variation of 0.1 to 2.0 magnitudes.} (see e.g. Ref. \cite{Pejcha:2011aa}). 

Following Ref. \cite{Riess:2016jrr} and breaking  the slope  of the Leavitt law at a period of $10$ days ($P=10\,d$) we include two different slopes $b_W^s$ and $b_W^l$ parameters in Eq. (\ref{wesmag}) for short and long period Cepheids with $P <10\, d$ and  $P >10\, d$ respectively (for a discussion about the presence of a broken PL slope at $P=10\,d$, see Refs. \cite{2002ASPC..283..258T,Tammann:2003ct,Sandage:2004qn,Ngeow:2005qc,Sandage:2008rq,2016MNRAS.457.1644B}). 

Ref.  \cite{Riess:2016jrr}  considers a universal reddening law in all galaxies and thus assume a global fixed value  $R_W=0.386$. This value is derived from the reddening law of Ref. \cite{Fitzpatrick:1998pb} using a ratio of total to selective extinction at the B and V bands\footnote{The total to selective extinction at B and V bands $R_V$ is defined as $R_V\equiv A_V/E(B-V)$} of the Johnson-Morgan or UBV (Ultraviolet, Blue, Visual) photometric system \cite{1963BAN....17..115J} $R_V=3.3$ to parameterize the shape of the extinction curve. In the literature the parameter $R_W$ ranges from $0.3$ to $0.5$ at H band (e.g.  $R_W=0.41$ in Ref. \cite{Riess:2011yx}) and the universal parameter $R_V$ ranges from $1$ to $6$ (the average value for the MW is $R_V = 3.1$) \cite{Krisciunas:2005nv,Nobili:2007zc,10.1093/mnras/stv881,Goobar:2008qp,Folatelli:2009nm,SDSS:2010swx,Goobar:2014noa,Amanullah:2014aba,Amanullah:2015rta,Cikota:2016htl,Biswas:2021qwe}  depending on the reddening law \cite{Cardelli:1989fp,1994ApJ...422..158O,Nataf:2015uma}. 

As noted recently by Ref. \cite{Mortsell:2021nzg} a global fixed value for parameter $R_W$ could result in a systematic error in distance measurements and the determination of the Hubble constant $H_0$. Refs. \cite{Mortsell:2021nzg,Mortsell:2021tcx} motivated by the observed variation in dust properties allowing for the parameter $R_W$ to vary between galaxies. However these studies make no attempt to search for possible transitions within the low z galaxy data. In the present analysis we  search for transition effects in Cepheid data at $z \lesssim 0.01$ ($\lesssim 40\,Mpc$).

\subsection{Datasets}
\label{data}

We use a sample of 1630 Cepheids in the anchor galaxies and in the  SnIa host galaxies. For 74 MW Cepheids including GAIA parallax measurements we use the dataset from Table 1 in Ref. \cite{Riess:2020fzl} and for 70 LMC Cepheids we use the dataset from Table 2 in Ref. \cite{Riess:2019cxk}.
These data are shown in Tables \ref{tab:mwcep} and \ref{tab:lmccep} of the Appendix \ref{DATA USED IN THE ANALYSIS}. Fitting the distance ladder with the system of equations for the 70 LMC Cepheids an intrinsic scatter of $0.08\, mag$ is added to the error estimates given in Table 2 of Ref. \cite{Riess:2019cxk} (note that Ref.  \cite{Riess:2019cxk} include a intrinsic LMC dispersion of $0.07 \,mag$) which is necessary in order to obtain a reduced chi-square of unity $\chi_{red}^2=\chi_{min}^2/dof=1$  \cite{Efstathiou:2020wxn}.

We obtain data for 1486 Cepheids in the anchor galaxy NGC 4258, in the M31 galaxy and in the 19 SnIa host galaxies from Table 4 in Ref \cite{Riess:2016jrr} (for details, see Ref. \cite{Hoffmann:2016nvl}). This data are shown in Table \ref{tab:hoscep} of the Appendix \ref{DATA USED IN THE ANALYSIS}. 

We consider the revised distance modulus to NGC 4258 of $\mu_{N4258}=29.397\pm 0.032 \,mag$  ($7.576 \pm 0.112 \, Mpc)$ reported by Ref. \cite{Reid:2019tiq} using water masers data. Also we consider the distance modulus to the LMC of $\mu_{LMC}=18.477\pm 0.0263\,mag$ derived by Ref. \cite{2019Natur.567..200P} with $1\%$ precision based on enhanced samples of late-type DEBs. This distance modulus is increased from that in Ref. \cite{Riess:2016jrr} by $0.5\%$.

Finally for SnIa B-band magnitudes we adopt the data from Table 5 in Ref. \cite{Riess:2016jrr}, derived  from the version 2.4 of SALT2\footnote{Spectral Adaptive Light Curve Template (SALT) is an empirical spectro-photometric model that is used in SnIa  light curve fitting \cite{Guy:2005me,SNLS:2007cqk,SNLS:2010pgl} (for SALT tests with simulations, see \cite{Mosher:2014gyd}). SALT is publicly available at \cite{SALT2.4}.} modeling of SnIa light curves by Ref. \cite{Betoule:2014frx}. These data are shown in Table \ref{tab:hossn} of the Appendix \ref{DATA USED IN THE ANALYSIS}.

Following Ref. \cite{Riess:2019cxk} for all LMC Cepheids we adopt the mean metallicity of $-0.33\,dex$  from Ref. \cite{Romaniello:2008yh} and $-0.27\,dex$  from Ref. \cite{10.1093/mnras/stv2414}, $[Fe/H]= -0.30\, dex$ which is slightly different than the value of $-0.25\, dex$ adopted by  Ref. \cite{Riess:2016jrr}. Also following Ref. \cite{Riess:2019cxk} we adopt $[O/H]=[Fe/H]$ (where O and Fe is the number of oxygen and iron atoms per unit of volume respectively) and we measure the metallicity in units of $Z=12+\log(O/H)$ (with Solar metallicity in these units $Z_{\odot}=12+\log(O/H)_{\odot}=8.824$ \cite{Mortsell:2021tcx}).

Note that since we use the datasets from Refs. \cite{Riess:2016jrr,Riess:2019cxk,Riess:2020fzl} with the same filters we do not need to apply further corrections.
The full datasets are available in \cite{githubfs}.

\section{SEARCH FOR TRANSITION}
\label{SEARCH FOR TRANSITION}

\begin{figure*} 
\begin{centering}
\includegraphics[width=0.87\textwidth]{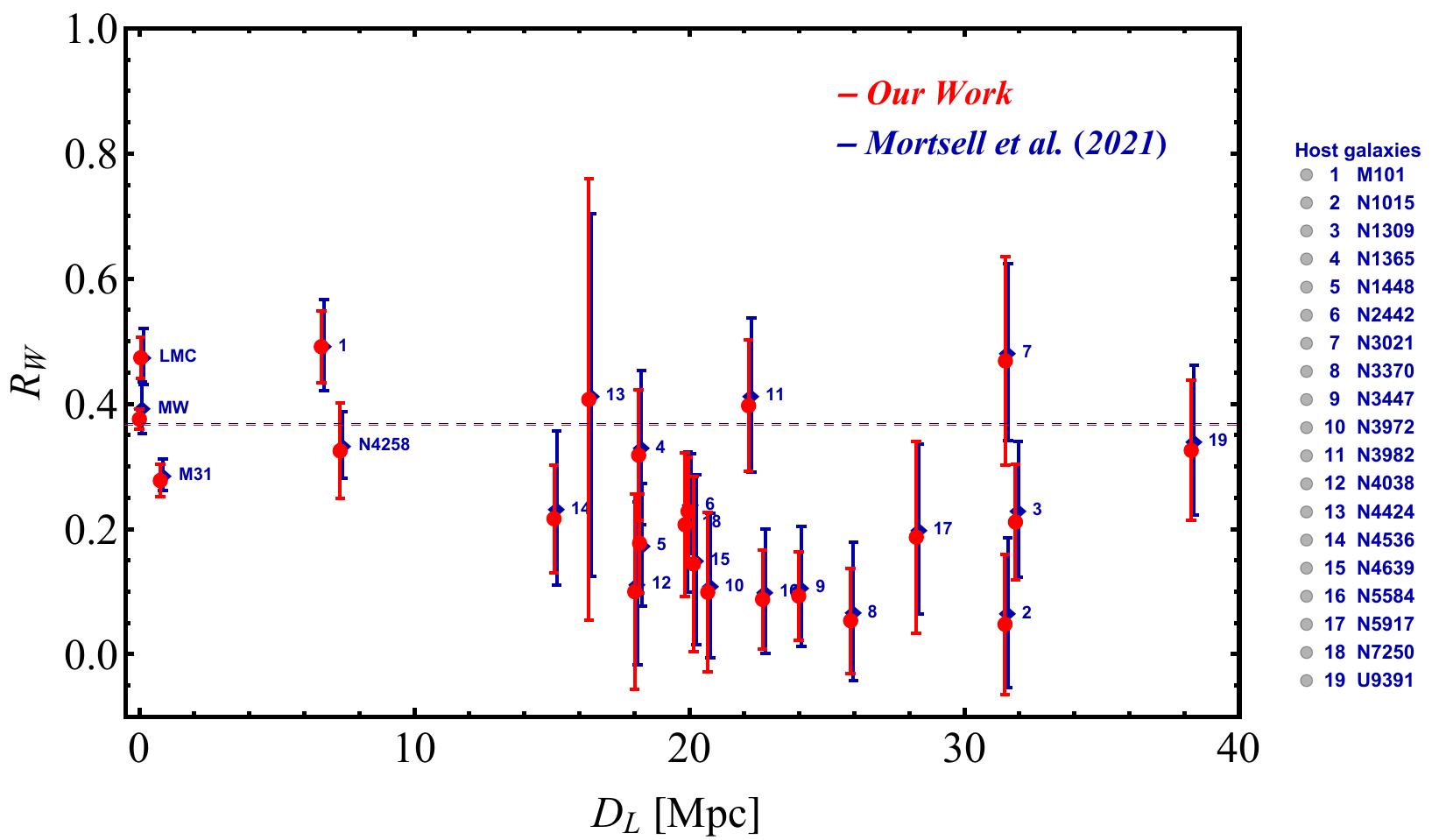}
\par\end{centering}
\caption{Fitting individual $R_W$ to Cepheid data as derived from our work (red points) and from Ref. \cite{Mortsell:2021nzg} (blue points). For  illustration purposes, the $D_L$ axis has been shifted slightly for our values so that the error bars do not overlap. The red and blue dotted lines correspond to $R_W=0.366$ and $R_W=0.369$ respectively.  These $R_W$ values are taken using the derived individual parameters of anchor galaxies and M31 (due to its proximity) $R_{W,k}$.} 
\label{fig4dl2} 
\end{figure*}

Our analysis is closely related to the results of study by Ref. \cite{Mortsell:2021nzg}. Using the same datasets that were used by Ref. \cite{Mortsell:2021nzg} we first reproduce their results and then we search for a transition signal of the  best fit parameter values $R_W$ and $M_H^W$ between data subsamples at low and high distances. 

From Eq. (\ref{wesmag}) using the Wesenheit magnitude of the $jth$ Cepheid in the $ith$ galaxy, including the host and the anchor galaxies, here MW, NGC 4258 and the LMC, separating PL relations for short and long period Cepheids we have
\begin{align}
&m_{H,i,j}^W\equiv m_{H,i,j}-R_{W,i}(V-I)_{i,j}\nonumber\\
&=\mu_i+M_{H,i}^W+b_W^s[P]_{i,j}^s+b_W^l[P]_{i,j}^l+Z_W[M/H]_{i,j}
\label{wesmagcep}
\end{align}
where $[P]_{i,j}^s=0$ for Cepheids with $P>10\,d$ and $[P]_{i,j}^l=0$ for Cepheids with $P<10\,d$.

In the case where $ith$ galaxy is the MW the distance modulus for the $jth$ Cepheid is estimated using parallaxes in units of mas (mas for milliarcsec) 
\be
\pi_j+zp=10^{-0.2(\mu_j-10)}
\label{parmw}
\ee
where $zp$ is a residual parallax calibration offset.

Thus
 \begin{align}
\mu_j&=10-\frac{5}{\ln{10}}\left[\ln{\pi_j}+\ln{\left(1+\frac{zp}{\pi_j}\right)}  \right]\nonumber\\
&\simeq 10-\frac{5}{\ln{10}}\left[\ln{\pi_j}+\frac{zp}{\pi_j}  \right]
\label{distmod}
\end{align}
where higher order terms $\mathcal{O}(zp/\pi_j)^2$ are negligible. 
Using Eq. (\ref{distmod}) into Eq. (\ref{wesmagcep}) in the case of MW Cepheids we obtain
\begin{align}
&m_{\pi,j}^W= m_{\pi,j}-R_{W}(V-I)_{j}\nonumber\\
&=M_{H}^W+b_W^s[P]_{j}^s+b_W^l[P]_{j}^l+Z_W[M/H]_{j}+\frac{5}{\ln{10}}\frac{zp}{\pi_j}
\label{wesmagcepmw}
\end{align}
where we use
\be
m_{\pi,j}^W=m_{H,j}^W-10+\frac{5}{\ln{10}}\ln{\pi_j}
\ee
and
\be
m_{\pi,j}=m_{H,j}-10+\frac{5}{\ln{10}}\ln{\pi_j}
\ee
Also in order to combine the measurements of SnIa and Cepheids we use the calibrated SnIa B-band peak magnitude in the $ith$ host 
\be
m_{B,i}=\mu_i+M_B
\label{wesmagsnia}
\ee

Using the data (see Section \ref{data}) for observed Cepheid magnitudes $m_H$, colors $V-I$, periods $[P]$, metallicities $[M/H]$, MW Cepheid parallaxes $\pi$, anchor distances $\mu_k$ together with the SnIa magnitudes $m_B$, we can fit simultaneously for $R_W$ , $b_W^s$, $b_W^l$, $Z_W$, the host and the anchor galaxy distances $\mu_i$, the parallax offset $zp$, the Cepheid absolute magnitude $M_H^W$ and the SnIa absolute magnitude $M_B$. 

Combining the equations for apparent magnitudes for Cepheids, Eqs. (\ref{wesmagcep}), (\ref{wesmagcepmw}), and for SnIa, Eq. (\ref{wesmagsnia}), we relate data and parameters through the matrix equation 
\be
\bf{Y} = \bf{AX }
\label{syst}
\ee
with $\bf{Y}$ the matrix of measurements, $\bf{X}$ the matrix of parameters and  $\bf{A}$ the equation (or design) matrix. Using these matrices with the measurement error matrix $\bf{C}$ we fit the data by minimizing the chi squared $\chi^2$ statistic expressed as 
\be
\chi^2=(\bf{Y}-\bf{AX})^T\bf{C}^{-1}(\bf{Y}-\bf{AX})
\label{chi2}
\ee
Note that we can solve the following expression for the maximum likelihood parameters $\bf{X}$  analytically: 
\be
\bf{X_{best}}=(\bf{A}^T\bf{C}^{-1}\bf{A})^{-1}\bf{A}^T\bf{C}^{-1}\bf{Y}
\label{bfpar}
\ee
The standard errors for the parameters in $\bf{X_{best}}$ are given by the covariance matrix
\be
\bf{\Sigma}=(\bf{A}^T\bf{C}^{-1}\bf{A})^{-1}
\label{covmat}
\ee

In the Appendix \ref{SYSTEM OF EQUATIONS} we present the schematic form of the  $\bf{C}$,  $\bf{Y}$, $\bf{X}$ and  $\bf{A}$  matrices. We adopt 2D fit including errors in the error matrix $\bf{C}$ in both $\bf{Y}$ and $\bf{X}$ "axes". In particular we do not neglect to include in the error matrix $\bf{C}$ the errors in the  colors V and I. These errors for MW and LMC Cepheids as provided by the SH$0$ES team (see in Table 1 of Ref. \cite{Riess:2020fzl} and in Table 2 of Ref. \cite{Riess:2019cxk}) are shown in Table \ref{tab:mwcep} and in Table \ref{tab:lmccep} of the Appendix \ref{DATA USED IN THE ANALYSIS} respectively. For Cepheids in galaxies (other than MW and LMC) where the SH$0$ES team does not provide separate color errors we use total statistical uncertainties where the color errors are included. These total statistical uncertainties derived by the SH$0$ES team (see in column 8 of the Table 4 of Ref. \cite{Riess:2016jrr}) are shown in Table \ref{tab:hoscep} of the Appendix \ref{DATA USED IN THE ANALYSIS}).

In the following subsections we study three cases where the questions mentioned in Introduction \ref{INTRODUCTION} will be addressed.

\subsection{Case I: Fitting individual $R_W$ and global $M_H^W$}

We first allow, as done in Ref. \cite{Mortsell:2021nzg}, the parameter $R_W$ to vary between galaxies and we consider a global value of the Cepheid absolute magnitude $M_H^W$. Despite slight differences in the analysis method, the results are in very good agreement with the results in Ref. \cite{Mortsell:2021nzg} as illustrated in Fig. \ref{fig4dl2}. Fitting individual parameters of galaxies $R_{W,i}$ to Cepheid data as derived from our work and from Ref. \cite{Mortsell:2021nzg} correspond to red and blue points respectively. The inferred best fit value of the Cepheid absolute magnitude is $M_H^W=-5.958\pm 0.028\,mag$. The red and blue dotted lines correspond to $R_W=0.365$ and $R_W=0.369$. These $R_W$ values are taken using the derived individual parameters of anchor galaxies (here MW, NGC 4258, and LMC) and M31 (due to its proximity) $R_{W,k}$ and minimizing the $\chi^2(R_W)$ with respect to the $R_W$ \be
\chi^2(R_W)=\sum_{k=1}^N\frac{(R_{W,k}-R_W)^2}{\sigma_{R_{W,k}}^2+\sigma_s^2}
\label{chirw}
\ee
where $N=4$. We fix the scatter to $\sigma_s = 0.08$ obtained by demanding that $\chi_{min}^2/N=1$ (where $\chi_{min}^2$ is the minimized value of $\chi^2$).

\begin{figure} 
\begin{centering}
\includegraphics[width=0.46\textwidth]{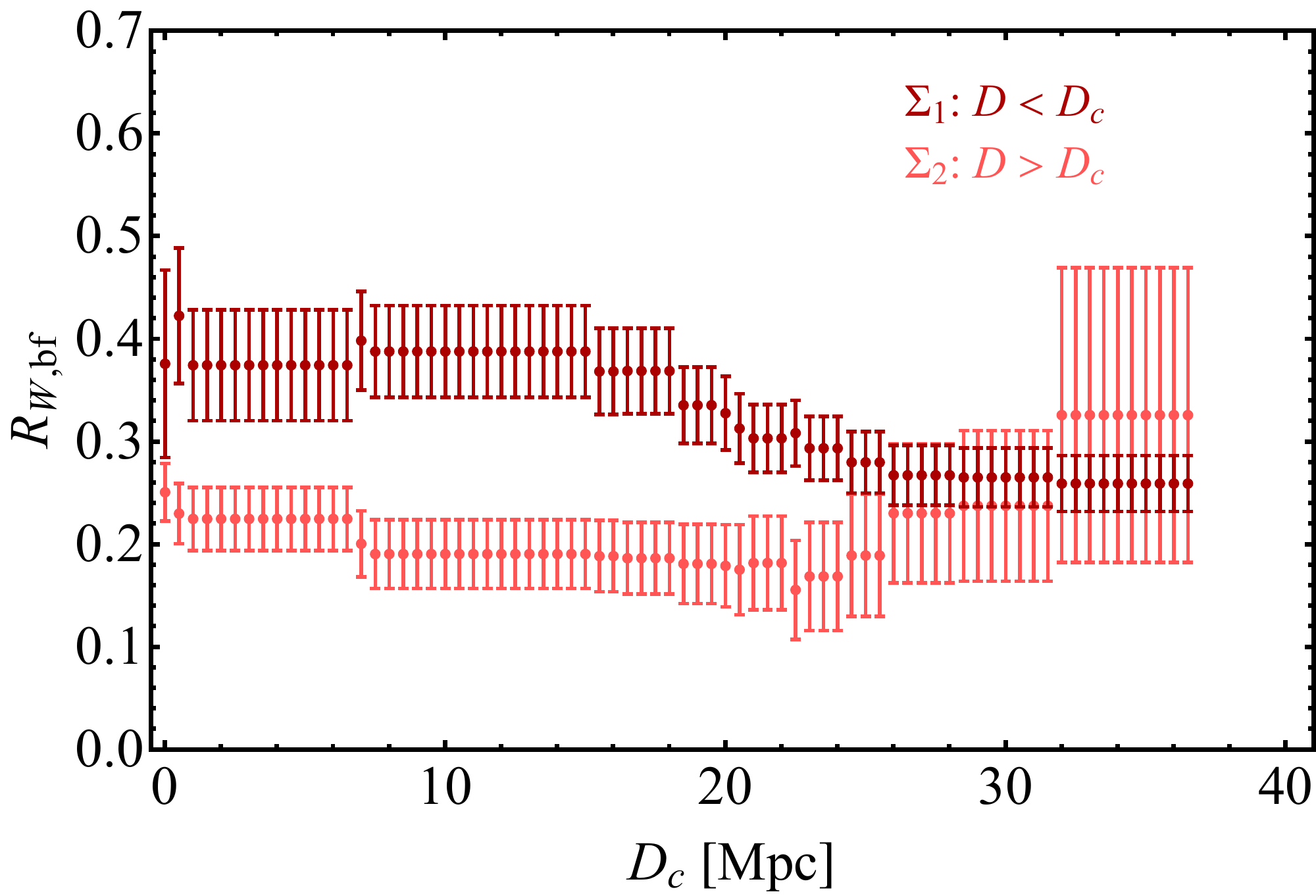}
\par\end{centering}
\caption{The best fit $R_{W,bf}$ for various $\Sigma_1$ and $\Sigma_2$ datasets as a function of the critical  dividing distance $D_c\in[0.01,37]\, Mpc$ as derived using the individual $R_W$ (red points in Fig. \ref{fig4dl2}). The dark red points correspond to the dataset with galaxies that have distance below $D_c$, whereas the red points regard galaxies with distances above $D_c$.} \label{figrw2sl} 
\end{figure}
\begin{figure} 
\begin{centering}
\includegraphics[width=0.46\textwidth]{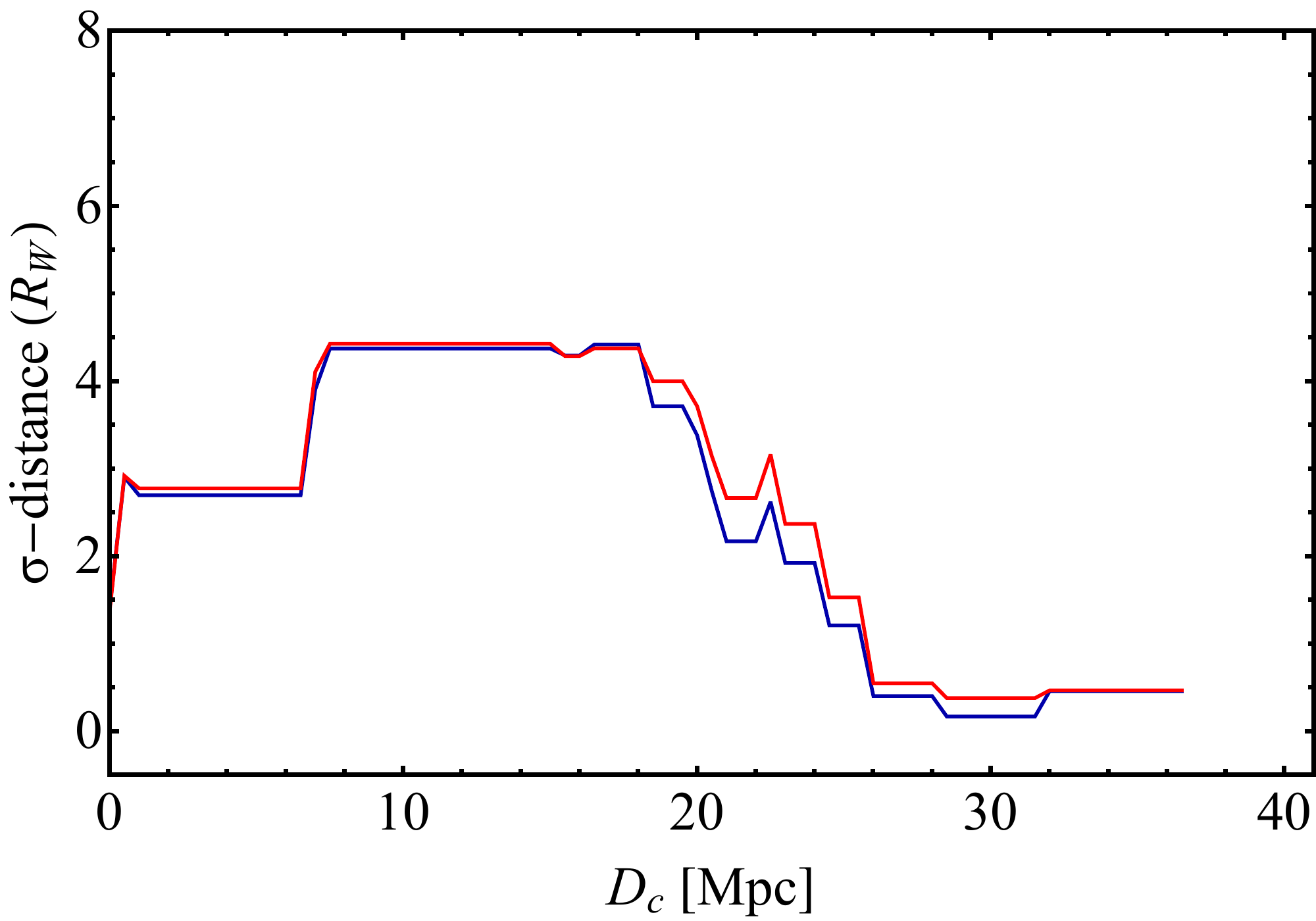}
\par\end{centering}
\caption{The $\sigma$-distances between the various  $\Sigma_1$ and $\Sigma_2$ datasets  as a function of the critical  dividing distance $D_c$ as derived using the individual values of $R_W$. The red and blue lines correspond to the red (our results) and blue (the results in Ref. \cite{Mortsell:2021nzg}) points of Fig. \ref{figrw2sl} respectively. A transition of the $\sigma$-distance at $D_c\simeq 22\, Mpc$ is apparent.} 
\label{figrwall} 
\end{figure}

Using the obtained best fit individual values for all galaxies $R_{W,i}$ (see red points in Fig. \ref{fig4dl2}) we focus on a particular type of evolution, sharp transition of these best fit values at low and high distances. We thus use the Distance Split Sample (DSS)  method which consists of the following steps:
\begin{itemize}
    \item 
We consider a critical dividing distance $D_c\in[0.01,37]\,Mpc$ and split the sample of galaxies in two subsamples $\Sigma_1$ and $\Sigma_2$ with distances $D < D_c$  and $D > D_c$ respectively. 
\item
For each subsample we use the maximum likelihood method to find the best fit parameters $R_{W,bf}$ ($R_W^<$ and $R_W^>$) by minimizing $\chi_1^2(R_W^<)$ and $\chi_2^2(R_W^>)$ using a similar equation as Eq. (\ref{chirw}). The best fit values  $R_{W,bf}$ for various $\Sigma_1$ and $\Sigma_2$ datasets as a function of the critical distances $D_c$ are shown in Fig. \ref{figrw2sl}.
\item
We evaluate the $\Delta \chi_{12}^2 (D_c)$ of the best fit of each subsample $\Sigma_1$ with respect to the likelihood of the other subsample $\Sigma_2$ and vice versa
\be
\Delta \chi_{12}^2(D_c)\equiv \chi_2^2(R_W^<)(D_c)-\chi_{2,min}^2(R_W^>)(D_c)
\ee
\be
\Delta \chi_{21}^2(D_c)\equiv \chi_1^2(R_W^>)(D_c)-\chi_{1,min}^2(R_W^<)(D_c)
\ee

\item
Using these values we evaluate the distances $d_{\sigma,12}(D_c)$ and $d_{\sigma,21}(D_c)$ as a solution of the equation
\be
\Delta \chi_{ij}^2=2 Q^{-1}\left[\frac{M}{2},1-Erf\left(\frac{d_{\sigma,ij}}{\sqrt{2}}\right)\right]
\label{dssol}
\ee
where $ij=12,\, 21$, $M$ is the number of parameters to fit i.e  $M=1$, $Q^{-1}$ is the inverse regularized incomplete Gamma function and $Erf$ is the error function.
\item
We then define the $\sigma-$distance $d_{\sigma}(D_c)$ as the minimum of the distances $d_{\sigma,12}(D_c)$ and $d_{\sigma,21}(D_c)$ i.e.
\be
d_{\sigma}(D_c)\equiv Min\left[d_{\sigma,12}(D_c),d_{\sigma,21}(D_c)\right]
\ee
The anticipated value of $d_{\sigma}$ is in the range of $1-2$ in the context of a homogeneous sample as verified below using Monte Carlo simulations. We thus address the question: \it{'Does the real Cepheid sample have this property?'}
\end{itemize}
\begin{figure*}
\begin{centering}
\includegraphics[width=0.9\textwidth]{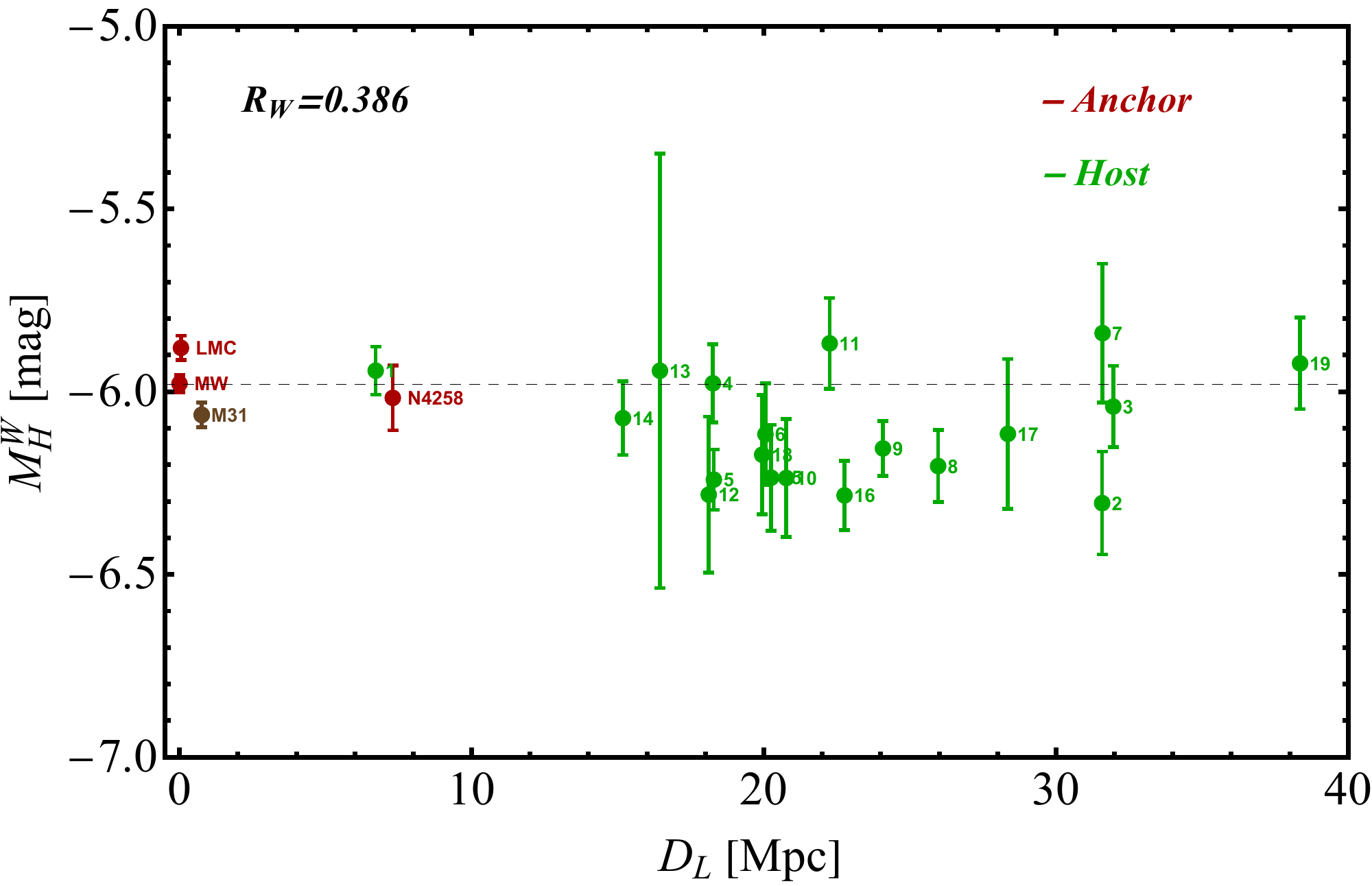}
\par\end{centering}
\caption{Fitting individual $M_H^W$ to Cepheid data for global $R_W$ with a fixed value $0.386$. Anchor galaxies are denoted with dark red points and SnIa host galaxies with green points. The dotted line corresponds to $M_H^W=-5.98\,mag$ as derived using the individual values of anchor galaxies and M31 (due to its proximity) $M_{H,k}^W$.} 
\label{figmhwdl} 
\end{figure*}
Fig. \ref{figrwall} shows the $\sigma-$distance $d_{\sigma}(D_c)$ between the various  $\Sigma_1$ and $\Sigma_2$ datasets as a function of the critical distances $D_c$ as derived using the individual values of $R_W$. The red and blue lines correspond to the red (our results) and blue (the results in Ref. \cite{Mortsell:2021nzg}) points of Fig. \ref{figrw2sl} respectively.

\begin{figure} 
\begin{centering}
\includegraphics[width=0.46\textwidth]{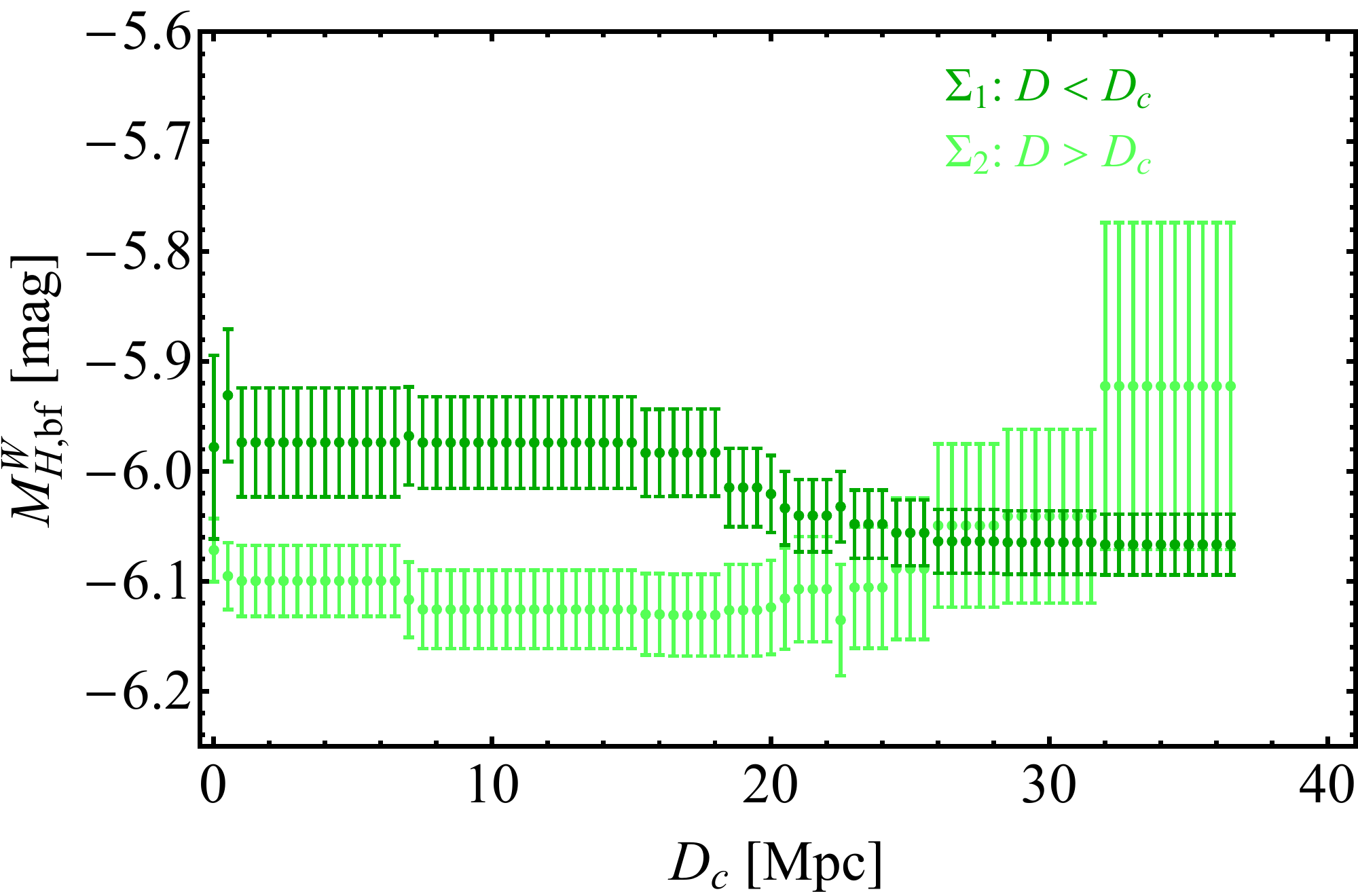}
\par\end{centering}
\caption{The best fit $M_{H,bf}^W$ for various $\Sigma_1$ and $\Sigma_2$ datasets as a function of the critical distances $D_c$ as derived using the individual values of $M_H^W$ (points in Fig. \ref{figmhwdl}). The dark green points correspond to the dataset with galaxies that have distance below $D_c$, whereas the green points regard galaxies with distances above $D_c$.} 
\label{figmhwsl} 
\end{figure}

Clearly, the Cepheid best fit $R_{W,bf}$ parameter indicates the presence of a transition at a critical distance $D_c$ in the range between $8Mpc$ and $18Mpc$ or at a time between $25\,Myrs$ and $55\,Myrs$ ago. For this range of $D_c$ the best fit value of $R_W^<= 0.388\pm 0.042$ differs from the best fit value of $R_W^>=0.206\pm 0.033$ at a level more that $4\sigma$ with $\Delta R_W\equiv R_W^> - R_W^<=-0.182\pm 0.056$. 

\subsection{Case II: Fitting individual $M_H^W$ with fixed global $R_W$}

In this case we allow the value of the Cepheid absolute magnitude $M_H^W$ to vary between galaxies and we consider a global fixed parameter $R_W=0.386$ in agreement with Refs. \cite{Riess:2016jrr,Riess:2019cxk,Riess:2020fzl}.

The results of fitting individual $M_H^W$ to Cepheid data are illustrated in Fig. \ref{figmhwdl}.  Anchor galaxies are denoted with dark red points and SnIa host galaxies with green points. We see that Cepheids in nearby galaxies are fainter. The dotted line corresponds to $M_H^W=-5.98\,mag$ as derived using the individual parameters of anchor galaxies (here MW, NGC 4258, and LMC) and M31 (due to its proximity) $M_{H,k}^W$ and minimizing $\chi^2(M_H^W)$ with respect to the $M_H^W$
\be
\chi^2(M_H^W)=\sum_{k=1}^N\frac{(M_{H,k}^W-M_H^W)^2}{\sigma_{M_{H,k}^W}^2+\sigma_s^2}
\label{chimhw}
\ee
where $N=4$. We fix the scatter to $\sigma_s = 0.08$ obtained by demanding that $\chi_{min}^2/N=1$.

\begin{figure} 
\begin{centering}
\includegraphics[width=0.46\textwidth]{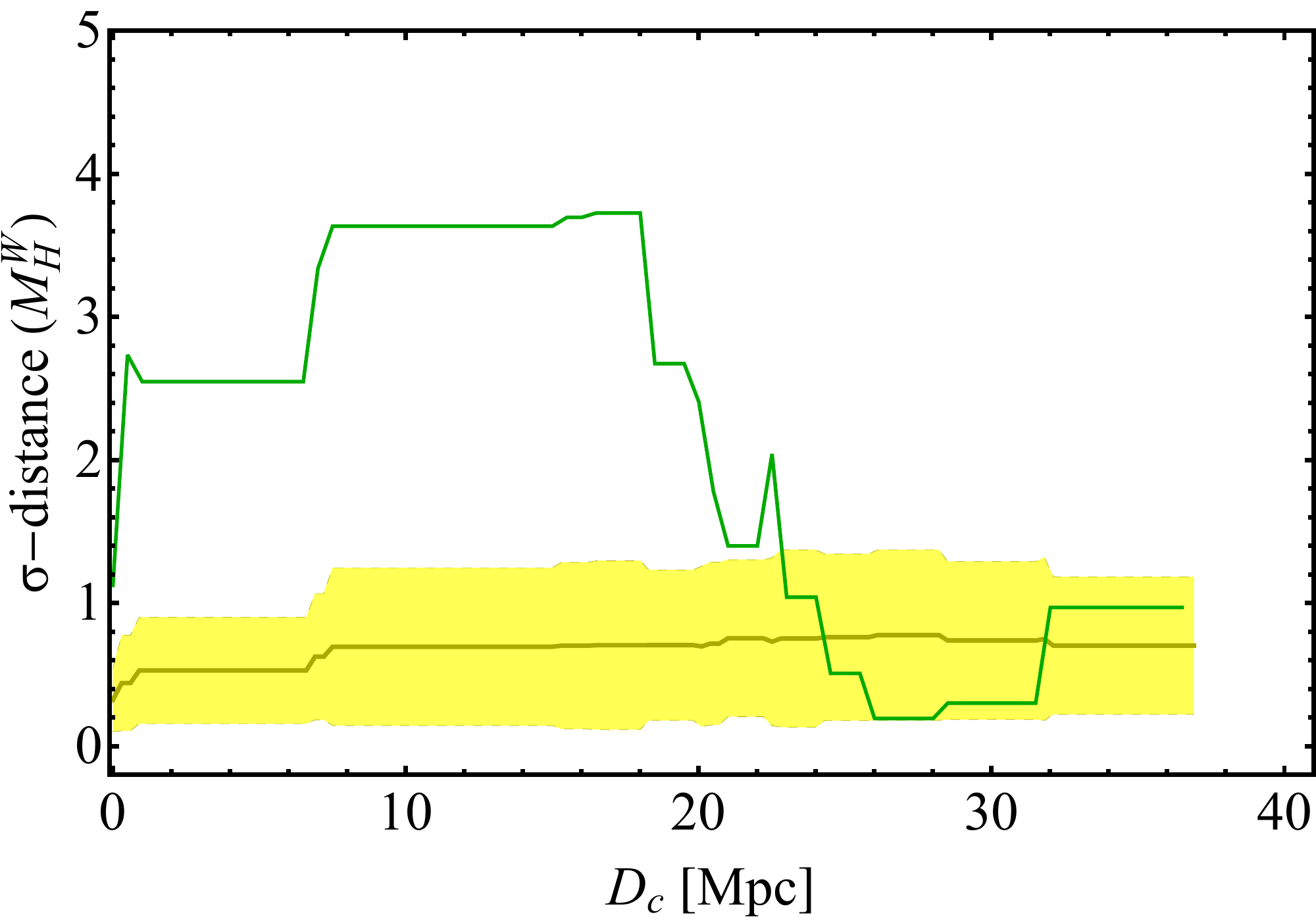}
\par\end{centering}
\caption{The green line represents the $\sigma$-distances between the various $\Sigma_1$ and $\Sigma_2$ datasets  as a function of the critical distances $D_c$ as derived using the individual values of $M_H^W$. In contrast the yellow lines correspond to $68\%$ (one standard deviation) range of the $\sigma$-distances as a function of the critical distances $D_c$ produced by a Monte Carlo simulation of 100 sample datasets assuming artificial homogeneity of the $M_H^W$ data. The simulations have been performed for randomly varying  $M_H^W$ values with a Gaussian probability distribution with  mean  $M_H^W=-6\,mag $ provided by the full $M_H^W$ datapoints and standard deviation  equal to the corresponding 1$\sigma$ error.} \label{figmhwnd1nofr} 
\end{figure}

Using the obtained best fit individual values for all galaxies $M_{H,i}^W$ (see Fig. \ref{figmhwdl}) we focus on a particular type of evolution, sharp transition of these best fit values at low and high distances. We follow  the same DSS method as in the previous subsection. Thus
\begin{itemize}
    \item 
First we consider a critical  dividing distance $D_c\in[0.01,37]\,Mpc$ and split the sample of galaxies in two subsamples $\Sigma_1$ and $\Sigma_2$ with distances $D < D_c$  and $D > D_c$ respectively.
\item
For each subsample we use the maximum likelihood method to find the best fit parameters $M_{H,bf}^W$ ($M_H^{W,<}$ and $M_H^{W,>}$) by minimizing $\chi_1^2(M_H^{W,<})$ and $\chi_2^2(M_H^{W,>})$. The best fit values of the $M_{H,bf}^W$ for various $\Sigma_1$ and $\Sigma_2$ datasets as a function of the critical distances $D_c$ are shown in Fig. \ref{figmhwsl}. 
\item
We consider the $\Delta \chi_{12}^2 (D_c)$ of the best fit of each subsample $\Sigma_1$ with respect to the likelihood of the other subsample $\Sigma_2$ and vice versa
\be
\Delta \chi_{12}^2(D_c)\equiv \chi_2^2(M_H^{W,<})(D_c)-\chi_{2,min}^2(M_H^{W,>})(D_c)
\ee
\be
\Delta \chi_{21}^2(D_c)\equiv \chi_1^2(M_H^{W,>})(D_c)-\chi_{1,min}^2(M_H^{W,<})(D_c)
\ee
\item
We evaluate the  distances $d_{\sigma,12}(D_c)$ and $d_{\sigma,21}(D_c)$ as a solution of the corresponding Eq. (\ref{dssol}). 
\item
We then find the $\sigma-$distances $d_{\sigma}(D_c)$ as the minimum of the distances $d_{\sigma,12}(D_c)$ and $d_{\sigma,21}(D_c)$. 
\end{itemize}
In Fig. \ref{figmhwnd1nofr} with the green line  we show the $\sigma-$distances $d_{\sigma}(D_c)$ between the various  $\Sigma_1$ and $\Sigma_2$ datasets as a function of the critical distances $D_c$ as derived using the individual values of $M_H^W$.
\begin{figure}
\begin{centering}
\includegraphics[width=0.46\textwidth]{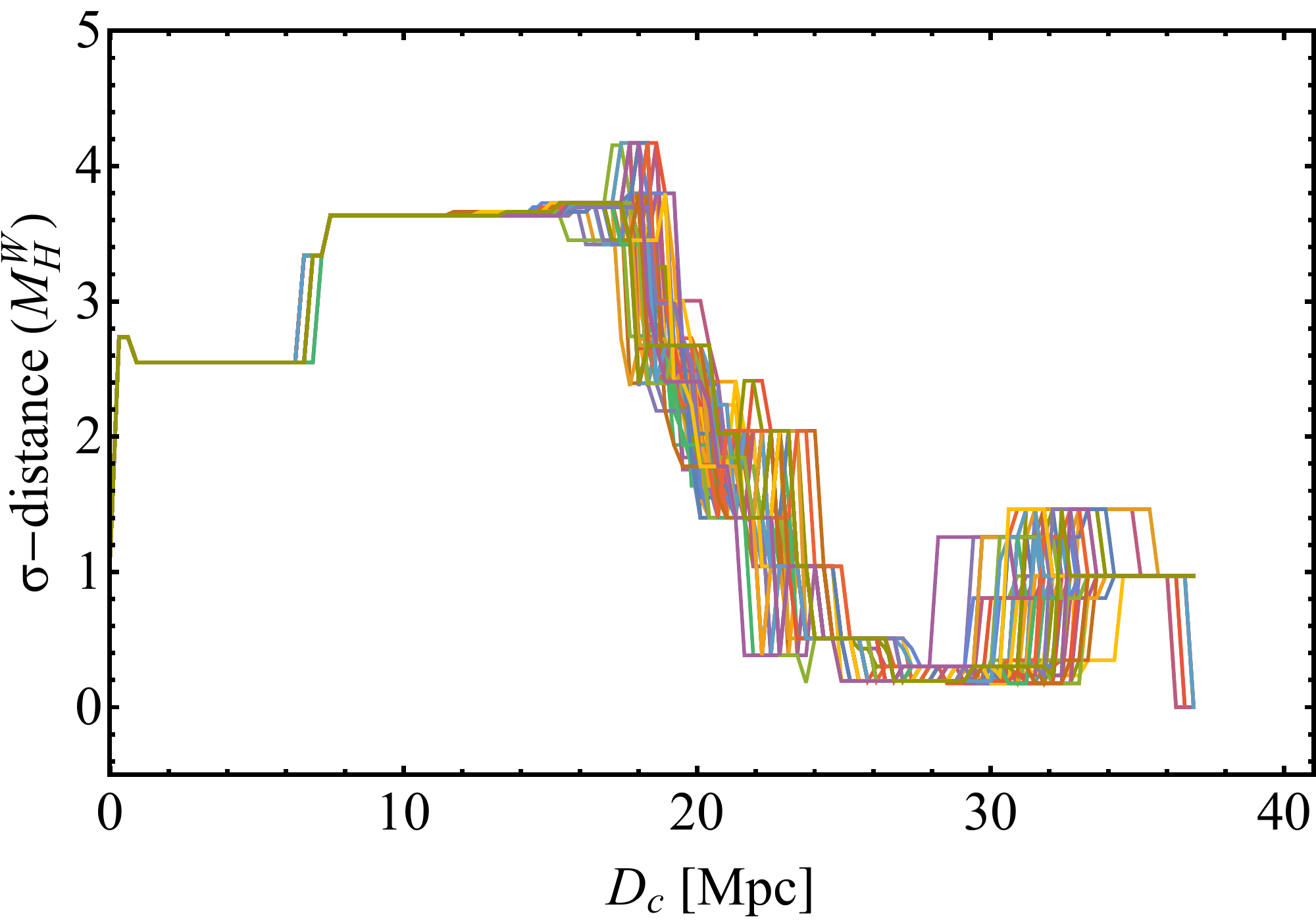}
\par\end{centering}
\caption{The  $\sigma$-distances as a function of the critical distances $D_c$ for 100 sample datasets with random distance values, normally distributed inside their individual 1-$\sigma$ range as derived using the individual values of $M_H^W$. A transition of the $\sigma$-distance at $D_c\simeq 22 Mpc$ remains present for practically all of the Monte Carlo samples.} 
\label{fig1mhwnd100} 
\end{figure}
\begin{figure}
\begin{centering}
\includegraphics[width=0.46\textwidth]{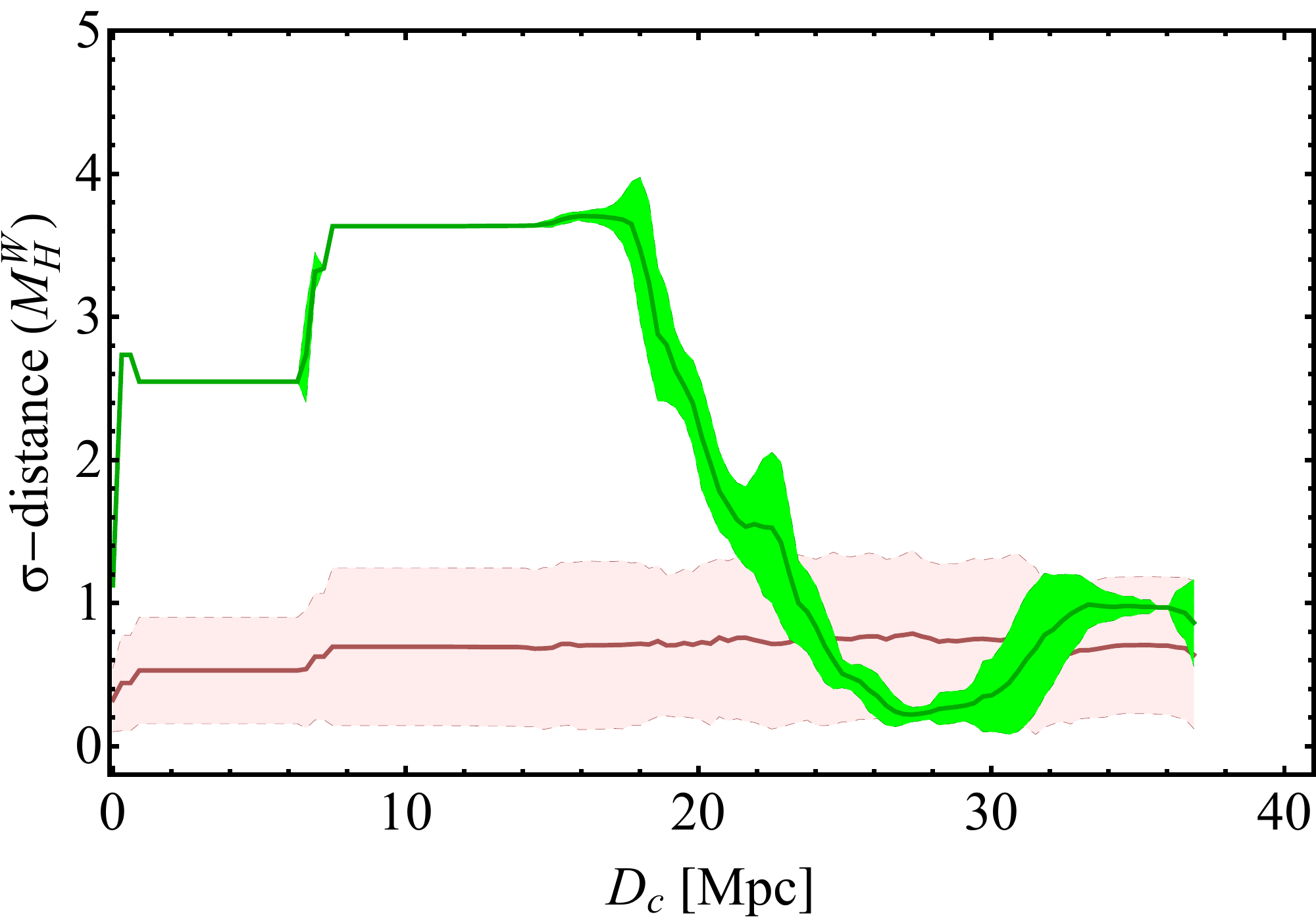}
\par\end{centering}
\caption{ The green lines represent the $68\%$  range of the $\sigma$-distances as a function of the critical distances $D_c$ produced by a Monte Carlo simulation of 100 sample datasets.  The simulations have been performed for randomly varying galaxy distance values with a Gaussian probability distribution with  mean  equal to the measured distance and standard deviation  equal to the corresponding 1$\sigma$ error. In contrast the pink region correspond to a Monte Carlo simulation of 100 sample datasets assuming artificial homogeneity of the $M_H^W$ data. In addition to this homogeneity the simulations have been performed for randomly varying galaxy distance values with a Gaussian probability distribution.} 
\label{figmhwnd2nofr} 
\end{figure}
\begin{figure*}
\begin{centering}
\includegraphics[width=1\textwidth]{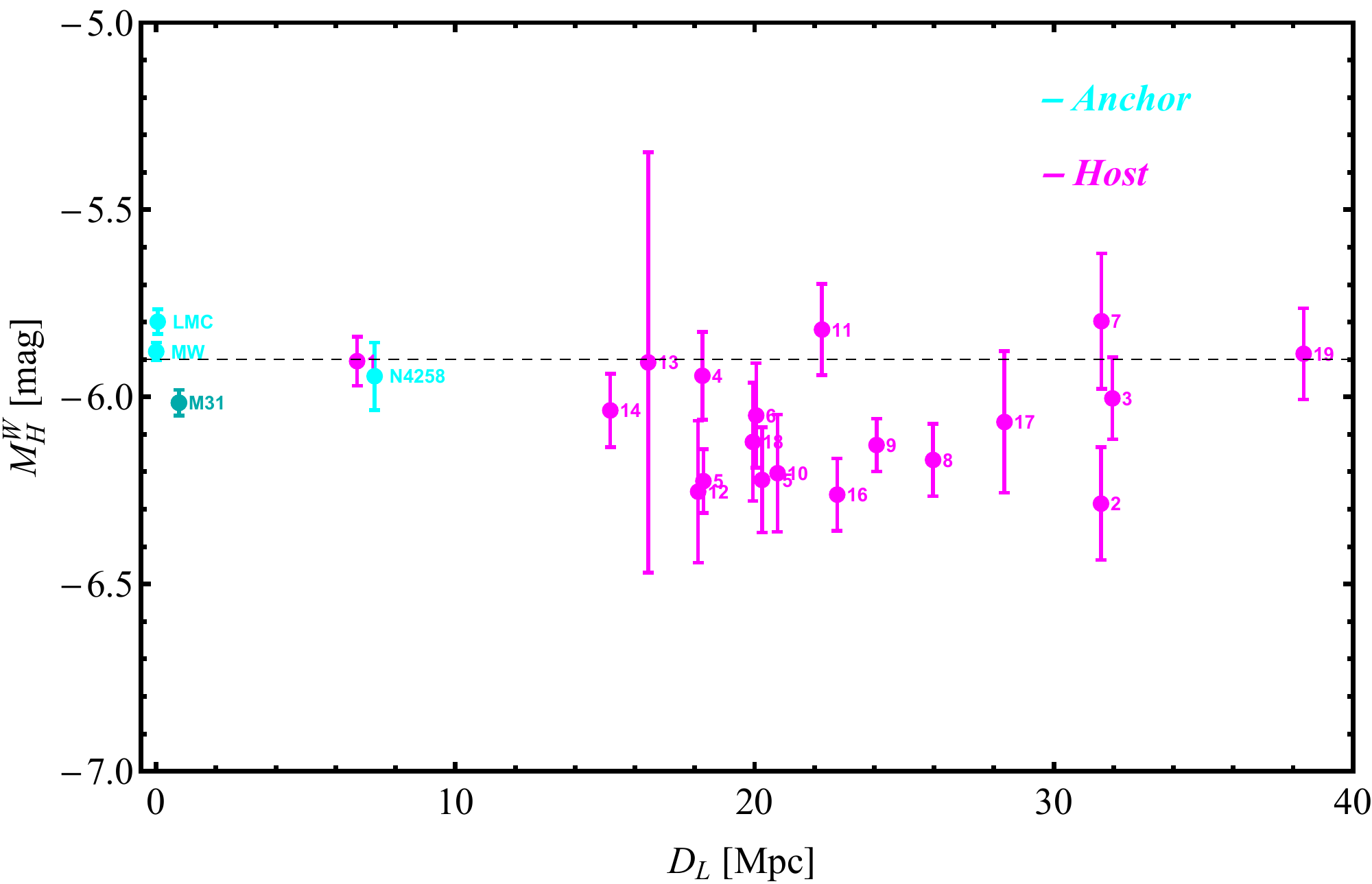}
\par\end{centering}
\caption{Fitting individual $M_H^W$ to Cepheid data with a free global $R_W$. Anchor galaxies are denoted with cyan points and SnIa host galaxies with magenta points. The dotted line corresponds to $M_H^W=-5.90\,mag$ as derived using the individual values of anchor galaxies and M31 (due to its proximity) $M_{H,k}^W$.} 
\label{figmhwdlrwfr} 
\end{figure*}
\begin{figure*}
\begin{centering}
\includegraphics[width=1\textwidth]{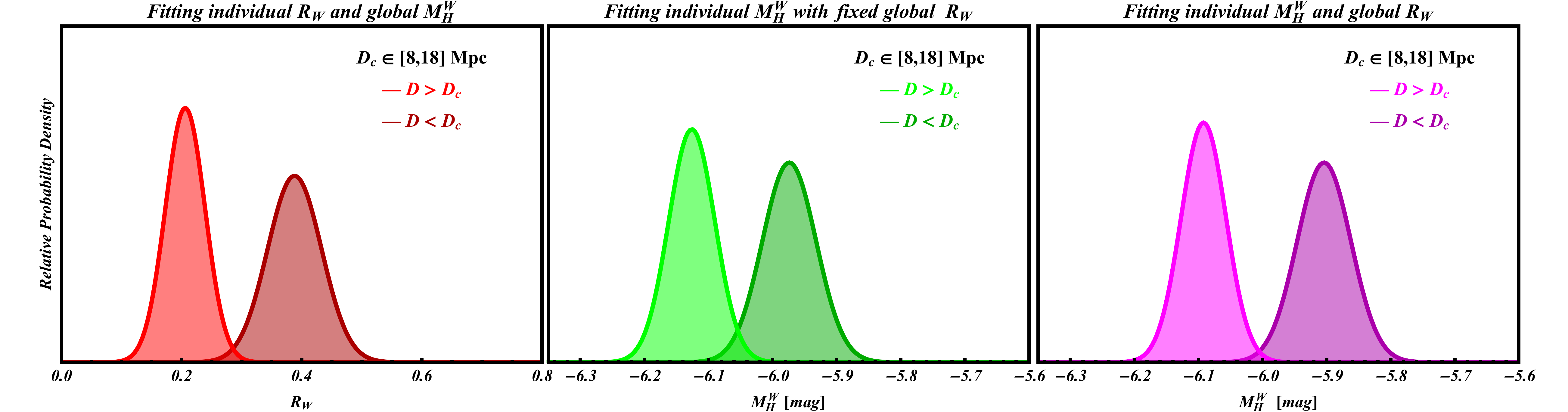}
\par\end{centering}
\caption{The one dimensional relative probability density values of the color luminosity parameter and the Cepheid absolute magnitude as derived using the DSS method for the cases I, II, and III. All measurements are shown as normalized Gaussian distributions. Notice that the best fit values one for galaxies at distances $D<D_c$ and  one for galaxies at $D>D_c$  are inconsistent with each other at a level larger than $3\sigma$.} 
\label{figprob} 
\end{figure*}
As in the previous case we see that the Cepheid best fit $M_{H,bf}^W$ parameter indicates the presence of a transition at a critical distance $D_c$ in the range between $8Mpc$ and $18Mpc$. For this range of $D_c$ the best fit value of $M_H^{W,<}= -5.974\pm 0.042\,mag$ differs from the best fit value of $M_H^{W,>}=-6.126\pm 0.036\,mag$  at a level more that $3\sigma$ with $\Delta M_H^W\equiv M_H^{W,>} - M_H^{W,<}=-0.152\pm 0.055\,mag$.

In order to secure the robustness of our approach we use a Monte Carlo simulation allowing the galactic distances to vary randomly using  their error bars. In particular, the simulations have been performed for randomly varying galaxy distance values with a Gaussian probability distribution (normal distribution) with mean equal to the measured distance and standard deviation equal to the corresponding 1$\sigma$ error. In Fig. \ref{fig1mhwnd100} we show the  $\sigma$-distances as a function of the critical distances $D_c$ for 100 sample datasets with random distance values, normally distributed inside their individual 1-$\sigma$ range as derived using the individual values of $M_H^W$. Clearly, the random variation of the galactic distances does not change the transition effect. The $68\%$ (one standard deviation) range of the $\sigma$-distances as a function of the critical distances $D_c$ produced by the Monte Carlo simulation of 100 sample datasets is shown in Fig. \ref{figmhwnd2nofr} with the green lines. Evidently, the Monte Carlo simulation demonstrates the robustness of the identified transition with respect to variation of galactic distances.

We now test if the effect would disappear in the context of homogenized Monte-Carlo Cepheid datasets. In Fig. \ref{figmhwnd1nofr}, we examine if this obvious transition of this case could be due to a systematic uncertainty of the Cepheid absolute magnitude value. Thus, the yellow region corresponds to the $68\%$ range of the $\sigma$-distances as a function of the critical distances $D_c$ produced by a Monte Carlo simulation of 100 sample datasets assuming artificial homogeneity of the $M_H^W$ data. The simulations have been performed by randomly varying the $M_H^W$ values with a Gaussian probability distribution with  mean  $M_H^W=-6\,mag$ as obtained by the full dataset with a universal $M_H^W$ and standard deviation of each global universal fit which is equal to the corresponding 1$\sigma$ error. For this Monte Carlo uniformized data there is no transition. This demonstrates that the observed transition is due to the actual Cepheid data and not to the method we used. As expected the same result persists if in addition to homogenizing the sample with respect to $M_H^W$  we also randomly vary the galactic distances as described above assuming a Gaussian distribution (pink region in Fig. \ref{figmhwnd2nofr}). 
Thus the observed transition effect as illustrated in Figs. \ref{figmhwnd1nofr} and \ref{figmhwnd2nofr} is robust with respect to random variation of the galactic distances and disappears only if we artificially homogenize the data in the context of Monte Carlo simulations.

\subsection{Case III: Fitted individual $M_H^W$ and a global $R_W$}

In this case we assume a free to fit global parameter $R_W$ and allow  the value of the Cepheid absolute magnitude $M_H^W$ to vary between galaxies. The results of fitting individual $M_H^W$ to Cepheid data are illustrated in Fig. \ref{figmhwdlrwfr}. The dotted line corresponds to $M_H^W=-5.90\,mag$ as derived using the individual parameters of anchor galaxies and M31 (due to its proximity) $M_{H,k}^W$ and minimizing  $\chi^2(M_H^W)$ in Eq. (\ref{chimhw}) with respect to the $M_H^W$. 

In this case we obtain the best fit value of the parameter $R_W=0.310\pm 0.021$ which is smaller than the  global fixed value $R_W=0.386$ used by Refs. \cite{Riess:2016jrr,Riess:2019cxk,Riess:2020fzl}. We attribute this difference to the fact that we have used the full Cepheid sample for its determination and not just the anchor galaxies and we have not used a global value of the absolute magnitude $M_H^W$ common for all Cepheids. 

Using the same DSS method as in the two previous cases we find the best fit values of the $M_{H,bf}^W$ ($M_H^{W,<}$ and $M_H^{W,>}$) for various $\Sigma_1$ and $\Sigma_2$ datasets as a function of the critical distances $D_c$. As the previous cases the presence of a transition at a critical distance $D_c$ in the range between $8Mpc$ and $18Mpc$ is obvious. For this range of $D_c$ the best fit value of $M_H^{W,<}= -5.904\pm 0.042\,mag$ differs from the best fit value of $M_H^{W,>}=-6.092\pm 0.035\,mag$  at a level more that $4\sigma$ with $\Delta M_H^W\equiv M_H^{W,>} - M_H^{W,<}=-0.188\pm 0.055\,mag$. 

In this case, we do not plot here the result since it is very similar to that plotted in the figures of previews cases but are available in our publicly available numerical analysis files \cite{githubfs}.

\begin{center}      
\begin{table*}     
\centering  
\begin{tabular}{c|c|c|cccc}  
\hhline{=======}
	&&    &&   &&      \\
 Model &Best Fit Parameters  (with 1$\sigma$ ranges) &  Model Selection  &&  \multicolumn{3}{c}{Intrinsic scatter of LMC Cepheids} \\
   & $\sigma_{LMC}=0.08$ ($\sigma_{LMC}=0$ )  & Criteria &&  $\sigma_{LMC}=0$   && $\sigma_{LMC}=0.08$     \\
 \hline 
{\bf Base-SH$0$ES } &$R_W=0.386 $ (fixed)& $\chi_{min}^2$   && 1767.48 && 1644.79    \\
   Global  $R_W$ &$M_H^W=-5.958\pm 0.028$& $\chi_{red}^2$  &&   1.089   &&  1.014      \\
 Global  $M_H^W$  & ${\bf M_B=-19.251\pm 0.057} $& $AIC$  &&  1823.48  &&   1700.79     \\
$N=1650$ &( $M_B=-19.261\pm 0.057$ )& $\Delta AIC$   &&  7.65 (0)  &&   17.97 (0)     \\
  $M=28$   &${\bf H_0=72.86\pm 1.95}$, \textcolor{ForestGreen}{${\bf H_0=73.50\pm 1.96}$} & $BIC$   && 1974.92  && 1852.23    \\
   $dof=1622$  &( $H_0=72.53 \pm 1.93$, \textcolor{ForestGreen}{$ H_0=73.17\pm 1.94$} )& $\Delta BIC$   &&   2.25 (0)   &&    12.57 (0)     \\  
    \hline 
{\bf Base}  &$R_W=0.309\pm 0.021$& $\chi_{min}^2$   && 1759.47  && 1624.82   \\
   Global  $R_W$ &$M_H^W=-5.862\pm 0.028$& $\chi_{red}^2$  &&   1.084   &&  1.002      \\
 Global  $M_H^W$  & ${\bf M_B=-19.225\pm 0.057} $& $AIC$  &&   1815.83  &&   1682.82     \\
$N=1650$ &( $M_B=-19.246\pm 0.054$ )& $\Delta AIC$   &&   0 (-7.65)  &&    0 (-17.97)    \\
  $M=29$   &${\bf H_0=73.73\pm 1.96}$, \textcolor{ForestGreen}{${\bf H_0=74.38\pm1.97 }$}& $BIC$   && 1972.67  && 1839.66     \\
   $dof=1621$  &( $H_0=73.03 \pm 1.86$, \textcolor{ForestGreen}{$ H_0=73.67\pm1.86 $} )& $\Delta BIC$   &&   0 (-2.25)   &&  0 (-12.57)     \\
 \hline
 &$R_{W,i}$ red points in Fig. \ref{fig4dl2}$\quad$&    &&      &&         \\
  {\bf  I}  	&$\quad$  $R_W^<=0.388\pm 0.045$ (using DSS)$\quad$& $\chi_{min}^2$  && 1676.76   && 1564.06    \\
 Individual $R_W$ &$\quad$ $R_W^>=0.206\pm 0.033$ (using DSS)$\quad$& $\chi_{red}^2$  && 1.049&& 0.978   \\
 Global $M_H^W$ &$M_H^W=-5.958\pm 0.028$& $AIC$ && 1778.76   && 1666.06     \\ 
$N=1650$  &${\bf M_B=-19.43\pm 0.056} $& $\Delta AIC$   && -37.07 (-44.72)     &&  -16.76 (-34.73)     \\
  $M=51$  &( $M_B=-19.491\pm 0.056$ )& $BIC$ && 2054.59 && 1941.9    \\ 
 $dof=1599$  &${\bf H_0=67.11\pm 1.76 }$, \textcolor{ForestGreen}{${\bf H_0=67.69\pm 1.77 }$} & $\Delta BIC$   &&  81.92 (79.67)    &&    102.24 (89.74)    \\
   	&( $H_0=65.24\pm 1.71$, \textcolor{ForestGreen}{$ H_0=65.81 \pm 1.72 $} ) &    &&    &&        \\
   \hline  
  	&$R_W=0.386$ (fixed)&    &&    &&        \\
{\bf   II }  &$M_{H,i}^W$  points in Fig. \ref{figmhwdl}$\quad$& $\chi_{min}^2$   &&1732.05 &&  1611.04       \\
  Global $R_W$& $\quad$ $M_H^{W,<}=-5.974\pm 0.042$ (using DSS)$\quad$ & $\chi_{red}^2$ && 1.083 && 1.007     \\
 Individual $M_H^W$  &$\quad$ $M_H^{W,>}=-6.126\pm 0.036$ (using DSS)$\quad$& $AIC$ 	&& 1832.05  &&  1711.04     \\
  $N=1650$ &${\bf M_B=-19.394 \pm 0.057} $& $\Delta AIC$   &&  16.22 (8.57)   &&  28.22  (10.25)     \\ 
     $M=50$  &( $M_B=-19.404\pm 0.055$ ) & $BIC$ &&2102.48  &&  1981.47     \\
 $dof=1600$   &${\bf H_0=68.22 \pm 1.82 } $, \textcolor{ForestGreen}{${\bf H_0=68.82 \pm 1.83 }$}& $\Delta BIC$   && 129.81  (127.56)   && 141.81  (129.24)      \\
   	& ( $H_0=67.90\pm 1.75 $, \textcolor{ForestGreen}{$ H_0=68.50\pm 1.76$} ) &    &&    &&        \\ 
   	\hline
 	&$R_W=0.310\pm 0.021$&    &&    &&        \\
{\bf   III}   & $M_{H,i}^W$ points in Fig. \ref{figmhwdlrwfr}$\quad$& $\chi_{min}^2$   &&  1726.7  &&  1592.09      \\
 Global $R_W$  &$\quad$ $M_H^{W,<}=-5.904\pm 0.042$ (using DSS)$\quad$& $\chi_{red}^2$  &&  1.079   && 0.996    \\
  Individual $M_H^W$ &$\quad$ $M_H^{W,<}=-6.092\pm 0.035$ (using DSS)$\quad$& $AIC$ 	&& 1828.7  && 1694.09    \\
  $N=1650$  &${\bf M_B=-19.428 \pm0.057} $& $\Delta AIC$   && 12.87 (5.22)    && 11.27   (-6.7)     \\
   $M=51$ &( $M_B=-19.424 \pm0.056$ )& $BIC$  && 2104.53 && 1969.93   \\ 
 $dof=1599$   &${\bf H_0=67.17\pm 1.79}$, \textcolor{ForestGreen}{${\bf H_0=67.76 \pm 1.80 }$}& $\Delta BIC$   && 131.86 (129.61)    && 130.27  (117.7)     \\
 	&( $H_0=67.28 \pm 1.75$, \textcolor{ForestGreen}{$ H_0=67.87 \pm 1.76 $} )&  &&   &&          \\
      \hline 
 {\bf   IV }  &$R_W^<=0.325\pm 0.018$ & $\chi_{min}^2$   &&  1744.19  &&  1611.65       \\
Two universal $R_W$     &$R_W^>=0.155\pm 0.054$& $\chi_{red}^2$  &&  1.077   &&  0.995     \\
 Global $M_H^W$  &$M_H^W=-5.885\pm 0.028$& $AIC$   &&   1804.19  &&   1671.46     \\
  $N=1650$  &${\bf M_B=-19.399 \pm0.057} $& $\Delta AIC$   &&  -13.34 (-18.99)   &&  -11.36  (-29.33)     \\
  $M=30$  &( $M_B=-19.447 \pm0.054$ )& $BIC$   && 1966.44  &&  1833.91     \\
 $dof=1620$   &${\bf H_0=68.06\pm 1.80}$, \textcolor{ForestGreen}{${\bf H_0=68.66\pm 1.81 }$} & $\Delta BIC$   && -6.23 (-8.48) &&  -5.75 (-18.32)     \\
 	&( $H_0= 66.59 \pm 1.66$, \textcolor{ForestGreen}{$ H_0=67.17\pm 1.67$} )&  &&   &&      \\
  \hline 
{\bf    V }  &$R_W=0.308\pm 0.021$& $\chi_{min}^2$   &&   1757.15  && 1621.98     \\
 Global $R_W$   &$M_H^{W,<}=-5.863\pm 0.024$& $\chi_{red}^2$  &&  1.085  &&  1.001      \\
Two universal $M_H^W$    &$M_H^{W,<}=-6.024\pm 0.062$& $AIC$   &&   1817.15  &&   1681.98  \\
  $N=1650$   &${\bf M_B= -19.361 \pm 0.057 }$& $\Delta AIC$   &&  1.32 (-6.33)   &&  -0.84  (-18.81)     \\
  $M=30$   &( $ M_B= -19.399 \pm 0.057$ )& $BIC$   &&   1979.41  && 1844.23     \\
$dof=1620$   &${\bf H_0= 69.27 \pm 1.82  }$, \textcolor{ForestGreen}{${\bf H_0=69.88\pm 1.83 }$}& $\Delta BIC$   &&  6.74 (4.49)   &&   4.57 (-8.0)     \\
   	&( $ H_0= 68.06 \pm 1.81 $, \textcolor{ForestGreen}{$ H_0=68.65\pm 1.82 $} ) &    &&    &&        \\
 \hline 
	&$R_W^<=0.329\pm 0.018$&  &&   &&       \\
{\bf    VI}   &$R_W^>=0.151\pm 0.053$& $\chi_{min}^2$   &&  1743.26   && 1612.09      \\
 Two universal $R_W$    &$M_H^{W,<}=-5.891\pm 0.024$& $\chi_{red}^2$  && 1.077   && 0.996     \\
  Two universal $M_H^W$  &$M_H^{W,>}=-5.900\pm 0.063$& $AIC$   &&  1805.26  &&   1674.09     \\
 $N=1650$  &${\bf M_B= -19.413 \pm 0.052 }$& $\Delta AIC$   &&  -10.57 (-18.22)   && -8.73  (-26.7)      \\
    $M=31$ &( $ M_B= -19.379 \pm 0.056$ )& $BIC$  &&  1972.93  &&  1841.75  \\
 $dof=1619$   &${\bf H_0= 67.62 \pm 1.64 }$, \textcolor{ForestGreen}{${\bf H_0=68.22\pm 1.65 }$}& $\Delta BIC$   &&  0.26 (-1.99)   &&  2.09   (-10.48)    \\
    	& ( $ H_0= 68.70 \pm 1.81 $, \textcolor{ForestGreen}{$ H_0=69.30\pm 1.81 $} )&    &&    &&        \\
       \hhline{=======}   
\end{tabular} 
\caption{Fitting results and model comparison tests for different models. For the $\Delta$AIC and $\Delta$BIC  the comparisons are made respect to base (base-SH$0$ES) models. The value of $H_0$ derived using the Eq. (\ref{ho1}) (black font) and the Eq. (\ref{ho2}) (green font). The best fit parameters of SnIa absolute magnitude $M_B$ and the value of $H_0$ in the parentheses correspond to intrinsic scatter of LMC Cepheids $\sigma_{LMC}=0$.}
\label{comp} 
\end{table*}   
\end{center}
\begin{figure*}
\begin{centering}
\includegraphics[width=1\textwidth]{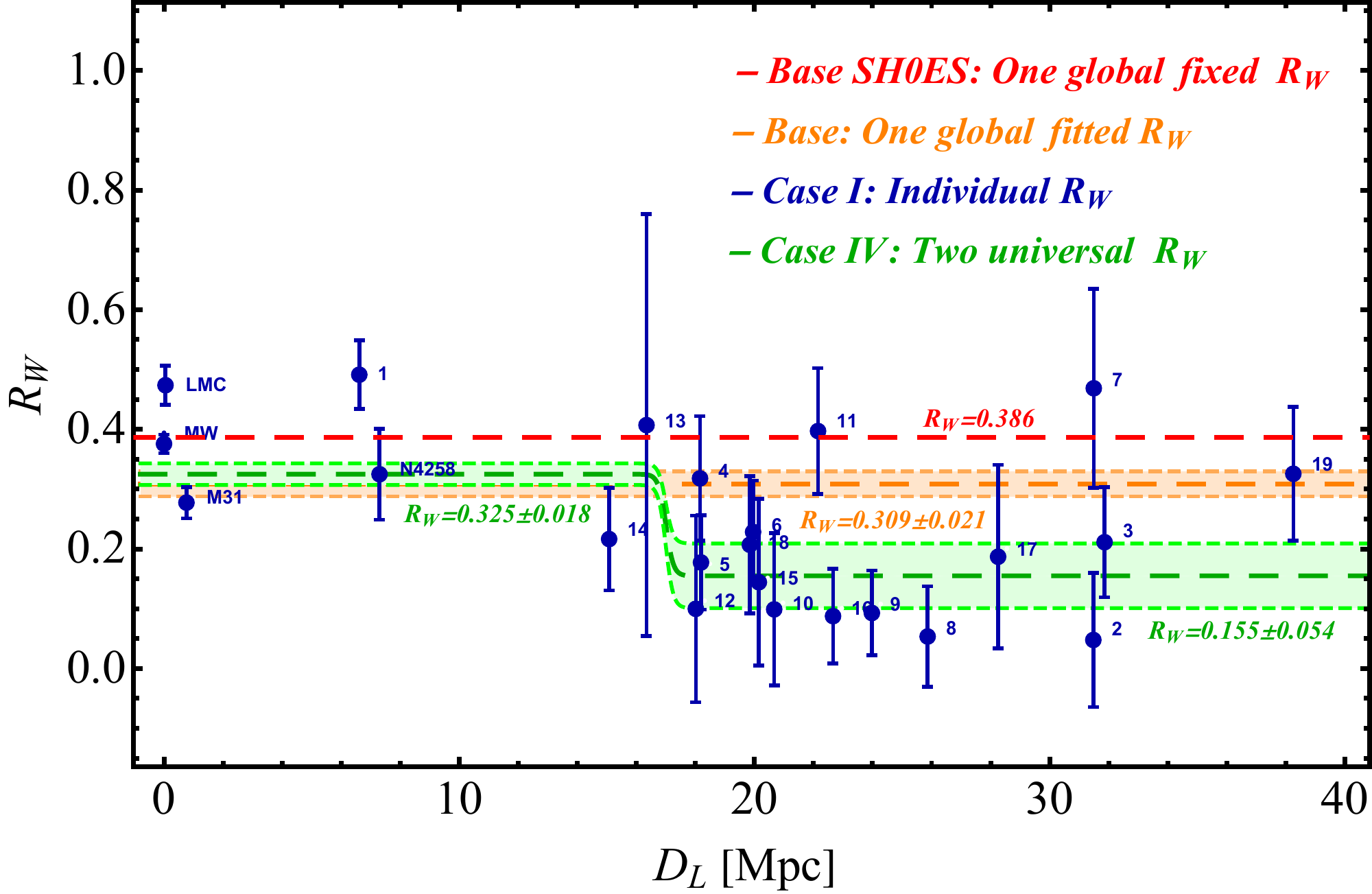}
\par\end{centering}
\caption{The best fit values of the parameter $R_W$ for base/base-SH$0$ES, I and IV models as derived using Cepheid data. Note that in terms of the AIC and BIC, fitting for two universal values of $R_W$ with global $M_H^W$ is the preferred model (case IV, green region).} 
\label{figtrans} 
\end{figure*}
\begin{figure*}
\begin{centering}
\includegraphics[width=1\textwidth]{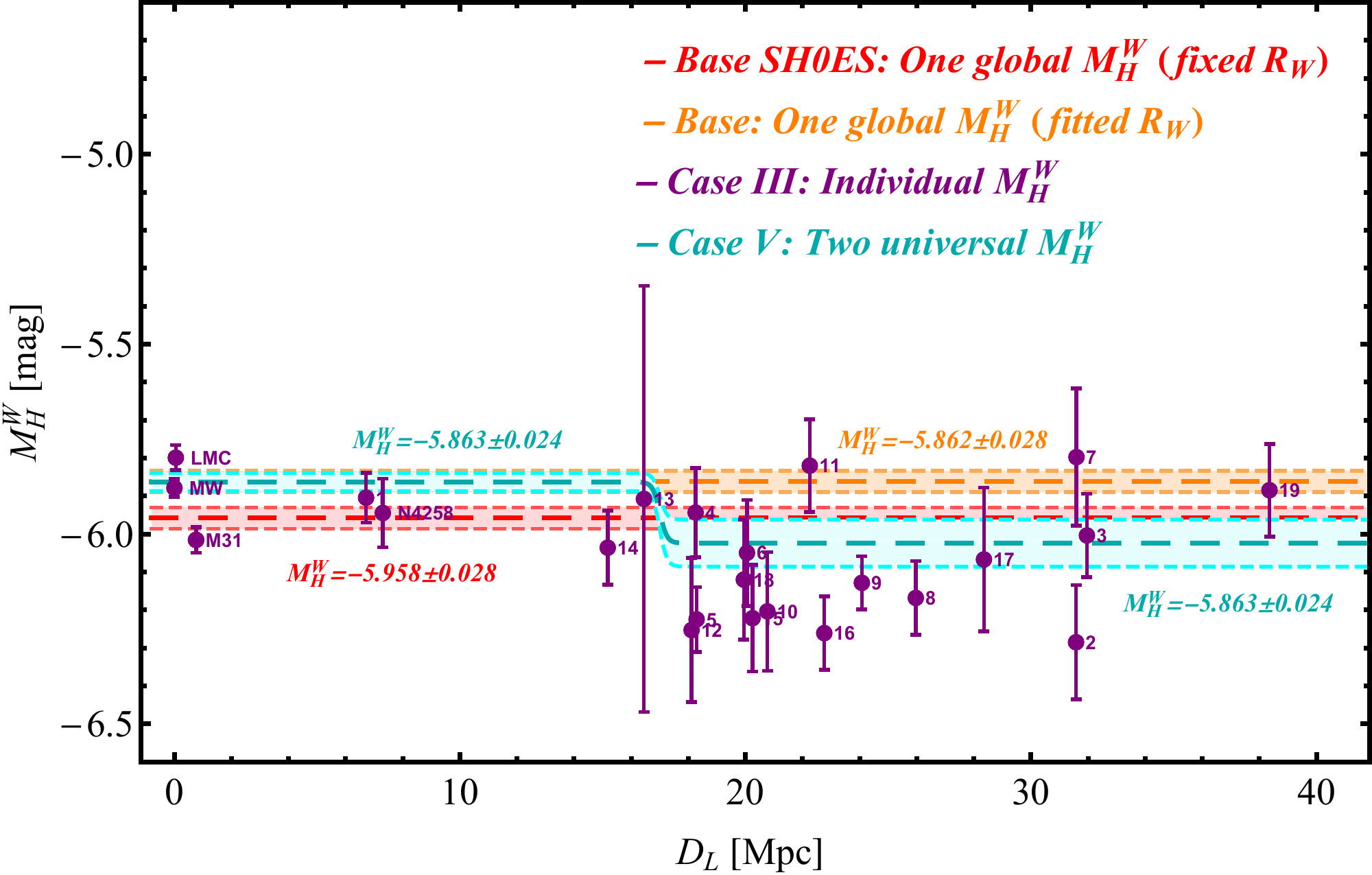}
\par\end{centering}
\caption{The best fit values of the parameter $M_H^W$ for base/base-SH$0$ES, III and V  models as derived using the Cepheid data. Note that in terms of the AIC and BIC, fitting for two universal values of $M_H^W$ with global $R_W$ is the preferred model among the models shown (case V, cyan region).} 
\label{figtransc} 
\end{figure*}

\begin{figure*}
\begin{centering}
\includegraphics[width=1\textwidth]{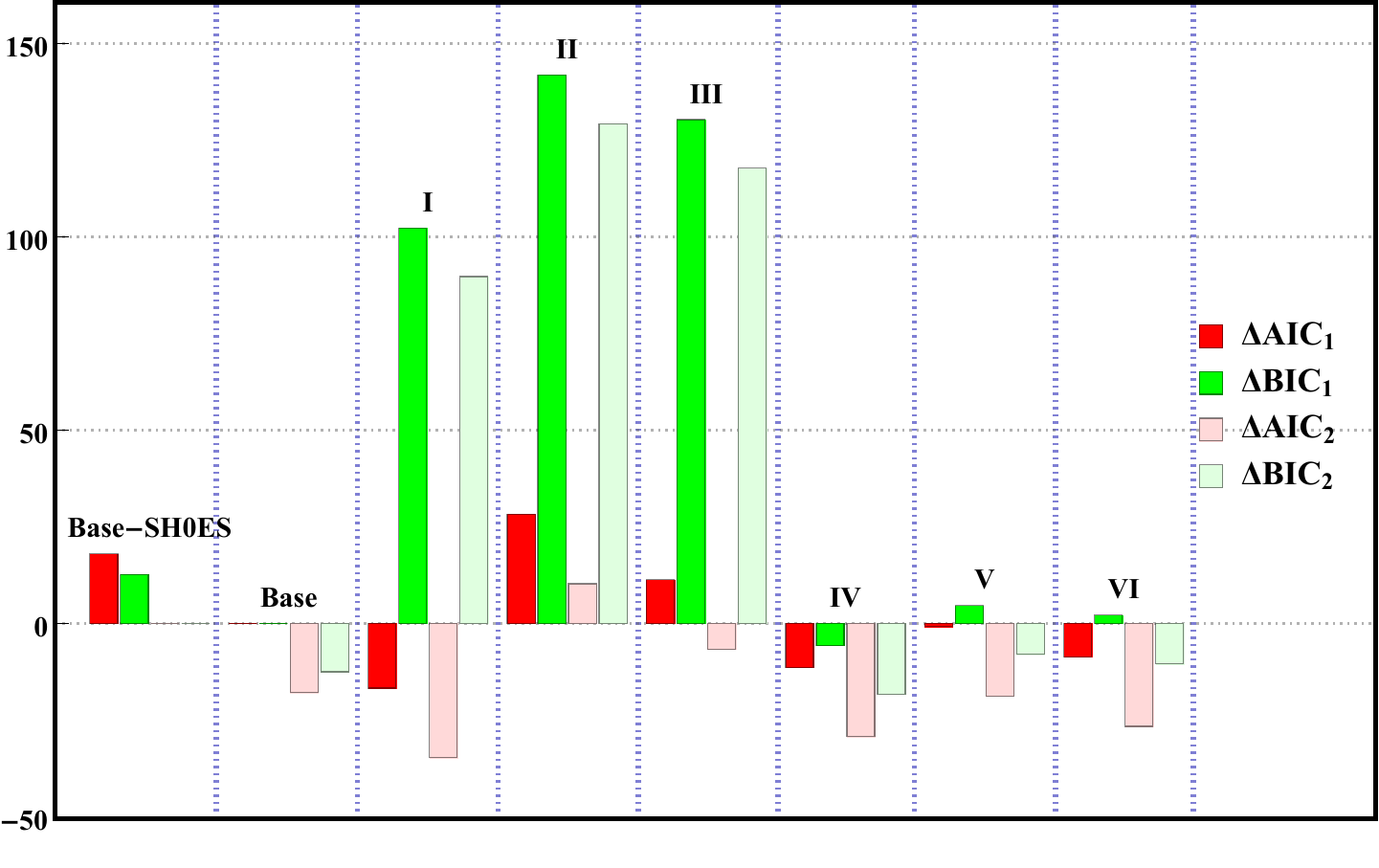}
\par\end{centering}
\caption{The $\Delta$AIC and $\Delta$BIC of models with  different free parameter set compared to base (subindex 1)  and base-SH$0$ES (subindex 2)  models. Clearly, the case IV (two universal $R_W$ and a global $M_H^W$) is the best model and on the other hand, the case II (a global $R_W$ and individual $M_H^W$ ) is the worst one.} 
\label{figbaric} 
\end{figure*}
\begin{center}      
\begin{table*}     
\centering  
\begin{tabular}{cccc|cccc}  
\hhline{========}
	&&    &&   &&   &   \\
\multicolumn{3}{c}{$\Delta AIC$ }  &&  \multicolumn{4}{c}{$\Delta BIC$ } \\
   &&     &&  &&    &   \\
    \hline 
    	&&    &&   &&   &  \\
\multicolumn{3}{c}{Level of empirical support for
the model  with the smaller $AIC$}  &&  \multicolumn{4}{c}{Evidence against the model
with the larger $BIC$} \\
   &&     &&  &&    &  \\
  \hline 
  &&     &&  &&    &  \\ 
  $\;$0-2  	&$\qquad$  4-7 &$ >10$   &&$\quad$ 0-2  &$\quad$2-6& $\quad$6-10  &$ >10$  \\ 
 
$\,\,$ Substantial$\;$ & $\qquad$ Strong & Very strong&& $\quad$Weak  &$\quad$Positive& $\quad$Strong &$\quad$Very strong  \\ 
 &&     &&  &&    &  \\
       \hhline{========}   
\end{tabular} 
\caption{The interpretation of differences $\Delta AIC$ and $\Delta BIC$ according to the calibrated Jeffreys’ scale \cite{Jeffreys:1961} (see also Refs. \cite{Liddle:2004nh,Nesseris:2012cq,BonillaRivera:2016use,Perez-Romero:2017njc,Camarena:2018nbr}). However, it should be noted that the Jeffreys’ scale has to be interpreted with care \cite{Nesseris:2012cq} because has been shown to lead to different qualitative conclusions.}
\label{jefsc} 
\end{table*}   
\end{center}

\begin{center}      
\begin{table*}     
\centering  
\begin{tabular}{c|ccc|ccc}  
\hhline{=======}
	&  &   &  &    \\
Ranking &\multicolumn{2}{c}{$ AIC$ } &&& \multicolumn{2}{c}{$BIC$ }\\
  	& $\sigma_{LMC}=0\:$ &  $\:\sigma_{LMC}=0.08$  &&& $\sigma_{LMC}=0\:$ &  $\:\sigma_{LMC}=0.08$  \\
	
    \hline 
	&  &   &  &  & &   \\    
  1	& I & I  &  &  &IV &IV  \\
2	& IV  & IV  &  &  &Base & Base \\
3	& VI & VI  &  &  & VI & VI \\
4	& Base & V  &  &  &Base-SH$0$ES & V \\
5	& V & Base  &  &  &V & Base-SH$0$ES \\
6	& Base-SH$0$ES & III   &  &  & I& I \\
7	& III & Base-SH$0$ES  &  &  &II &III  \\
8	& II & II  &  &  &III & II \\
	&  &   &  &  & &   \\	
       \hhline{=======}   
\end{tabular} 
\caption{Ranking of models according to $AIC$ and $BIC$ criteria. We see that in terms of the AIC and BIC fitting for two universal values of $R_W$ with global $M_H^W$ is the preferred model (case IV).}
\label{rank} 
\end{table*}   
\end{center}

\section{Model Selection}
\label{selection}
As pointed out in the previous subsections where we study three cases (I, II, and III) both the Cepheid best fit absolute magnitude and color luminosity parameters indicate the presence of a transition effect at a critical distance $D_c$ in  the  range  between  $8\,Mpc$ and  $18\,Mpc$ (see Fig. \ref{figprob}). This transition however becomes apparent when additional parameters are introduced (the individual $R_W$ or $M_H^W$ for each galaxy). Thus the questions we address in this section is the following:
\begin{itemize}
\item
Is the introduction of additional parameters favored by model selection criteria like the Akaike Information Criterion (AIC) \cite{akaike1974new} and the Bayesian Information Criterion (BIC) \cite{Schwarz:1978tpv}? 
\item
Could the introduction of a smaller number of parameters lead to more favored phenomenological models?
\end{itemize}

In order to address these questions we use model selection tests for eight cases with different number of
free parameters. The additional five considered cases are the following\footnote{Note that for all cases we fit other 27 additional parameters (see the schematic form of the matrix of parameters {\bf X} in Appendix \ref{SYSTEM OF EQUATIONS}) }:
\begin{itemize}
\item 
    {\bf Base-SH$0$ES}:  Like  previous studies of SH$0$ES team we consider universality on the color-luminosity relation with a global fixed parameter $R_W=0.386$ \cite{Riess:2016jrr,Riess:2019cxk,Riess:2020fzl} and universality on the absolute magnitude of Cepheids SnIa calibrators with a global $M_H^W$ to be fit by the Cepheid data. Thus in this case we use the base, commonly used parameter set in the field.
    \item 
    {\bf Base}: We consider universality on the color-luminosity relation with a global parameter $R_W$ to be fit by data and universality on the absolute magnitude of Cepheids SnIa calibrators with global $M_H^W$.
    \item 
    {\bf IV}: We consider a global $M_H^W$ and two universal $R_W$ ($R_W^<$ for galaxies  at  distances $D < 16\,Mpc$ and $R_W^>$ for galaxies  at  distances $D > 16\,Mpc$).
    \item 
    {\bf V}: We consider a global $R_W$ and two universal $M_H^W$ ($M_H^{W,<}$ for galaxies  at  distances $D < 16\,Mpc$ and $M_H^{W,>}$ for galaxies  at  distances $D > 16\,Mpc$).
     \item 
    {\bf VI}: We consider two universal $R_W$ ($R_W^<$ for galaxies  at  distances $D < 16\,Mpc$ and $R_W^>$ for galaxies  at  distances $D > 16\,Mpc$) and two universal $M_H^W$ ($M_H^{W,<}$ for galaxies  at  distances $D < 16\,Mpc$ and $M_H^{W,>}$ for galaxies  at  distances $D > 16\,Mpc$).
\end{itemize}
In order to compare the models we construct Table \ref{comp} with the fitting parameters for all cases. The best fit values of the parameter $R_W$ for base/base-SH$0$ES, I and IV models are shown in Fig. \ref{figtrans}  and  the best fit values of the parameter $M_H^W$ for base/base-SH$0$ES, III  and V  models are shown in Fig. \ref{figtransc}.
\begin{figure*}
\begin{centering}
\includegraphics[width=1\textwidth]{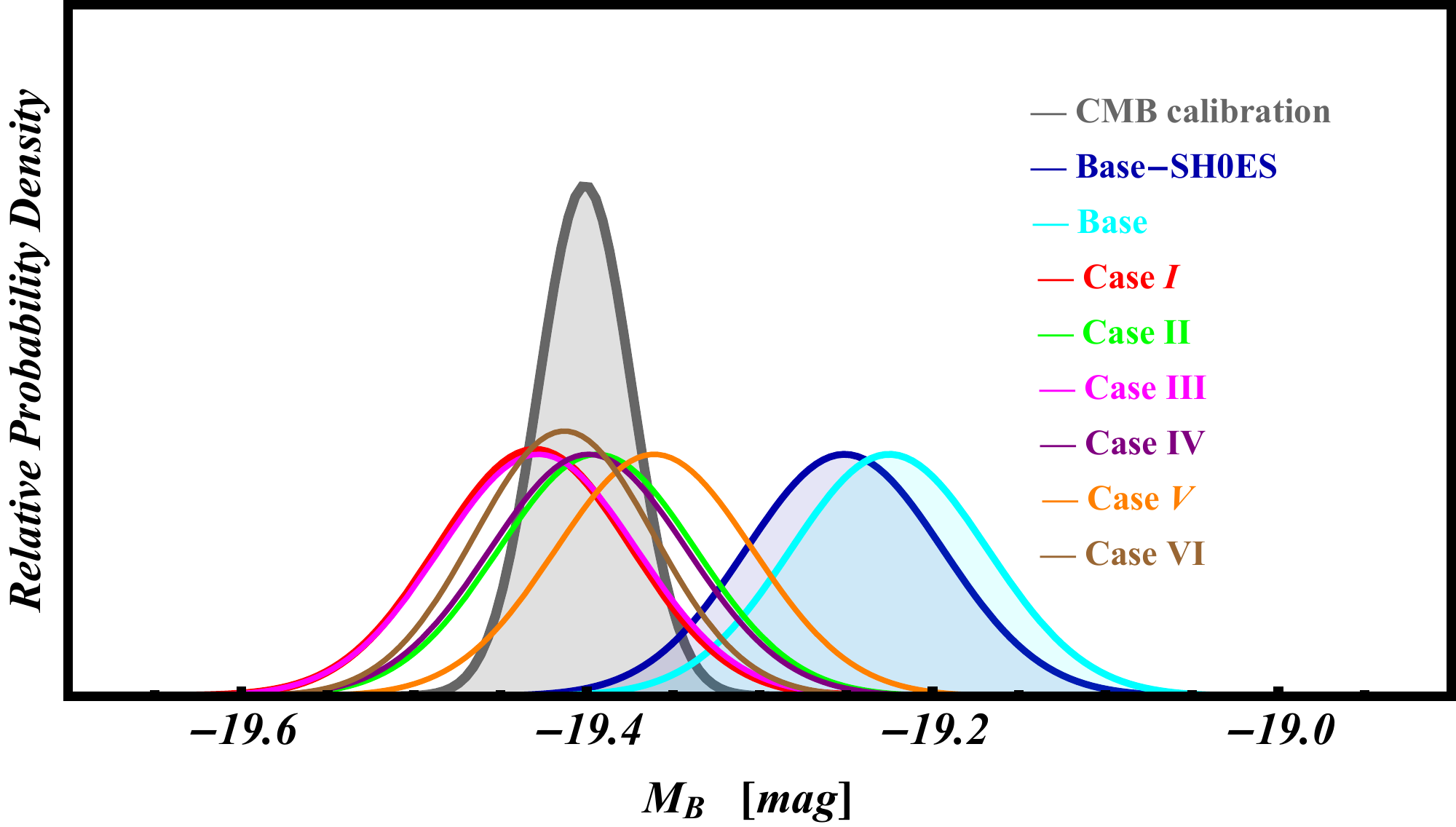}
\par\end{centering}
\caption{The one dimensional relative probability density value  of SnIa absolute magnitude $M_B$ for all cases studied in this paper compared to that obtained using CMB calibration. All measurements are shown as normalized Gaussian distributions. Clearly, for all cases where we do not consider the universality of parameters $R_W$ and $M_H^W$ (i.e. I, II, III, IV, V, VI) the $M_B$ is consistent with the CMB determination value.} 
\label{figprmb} 
\end{figure*}
\begin{figure*}
\begin{centering}
\includegraphics[width=1\textwidth]{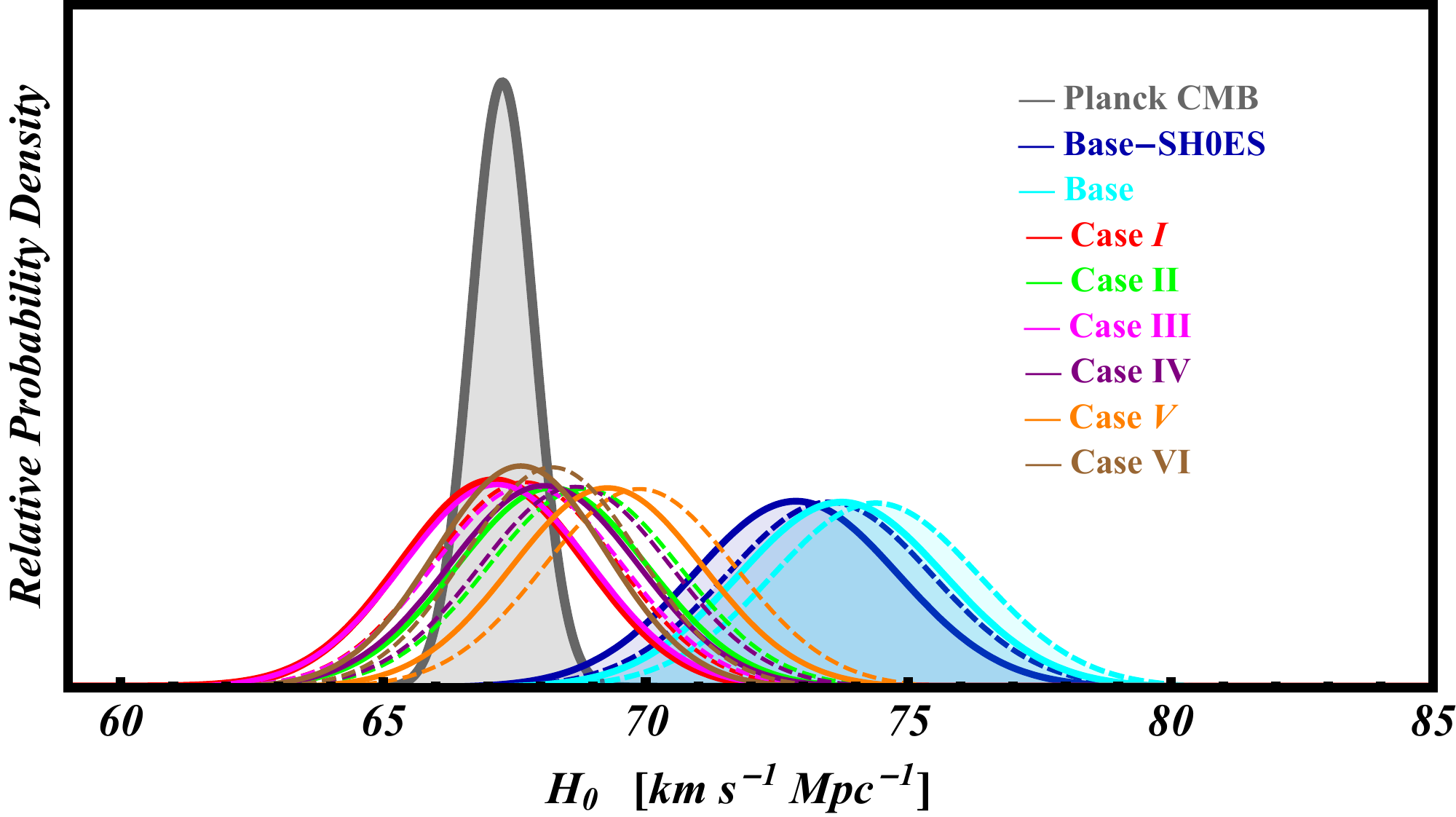}
\par\end{centering}
\caption{The one dimensional relative probability density value  of $H_0$ as derived using the Eq. (\ref{ho1}) (solid lines) and the Eq. (\ref{ho2}) (dashed lines) for all cases  studied in this paper compared to that from the Planck CMB measurement (grey line). All measurements are shown as normalized Gaussian distributions. It is evident that for all cases where we break the assumption of universality of the parameters $R_W$ and $M_H^W$ (i.e. I, II, III, IV, V, VI) the derived values of $H_0$ are consistent with the corresponding predicted Planck CMB best fit value.} 
\label{figprh0} 
\end{figure*}
Various methods for model selection have been developed  and model comparison techniques used  \cite{Liddle:2004nh,Liddle:2007fy,Arevalo:2016epc,Kerscher:2019pzk}. In table \ref{comp} we show the value of the minimum $\chi^2$ ($\chi_{min}^2$) for all cases and the reduced chi-squared  which is a very popular method for model comparison. This is defined by 
\be
\chi_{red}^2=\frac{\chi_{min}^2}{dof}
\ee
where $dof=N-M$ is typically the number of degrees of freedom (with $N$ is the number of datapoints used in the fit and $M$ is the number of free  parameters) for each model.

We also use the model selection methods like  Akaike Information Criterion (AIC) \cite{akaike1974new} and the Bayesian Information Criterion (BIC) \cite{Schwarz:1978tpv} that penalize models with additional parameters. For a model with $M$ parameters and a dataset with $N$ total observations these are defined through the relations  \cite{Liddle:2004nh,Liddle:2007fy,Arevalo:2016epc}
\be 
AIC=-2ln\mathcal{L}_{max}+2M=\chi_{min}^2+2M
\label{aic}
\ee
\be 
BIC=-2ln\mathcal{L}_{max}+Mln{N}=\chi_{min}^2+Mln{N}
\label{bic}
\ee 
where $\mathcal{L}_{max}\equiv e^{-\chi_{min}^2/2}$  (e.g. \cite{John:2002gg,Nesseris:2012cq}) is the maximum likelihood of the model under consideration.  Note that a version of the AIC corrected for small sample sizes is important \cite{Burnham:2002,Burnham:2004}. This version is given by \cite{doi:10.1080/03610927808827599} (see also Refs. \cite{Liddle:2007fy,Arjona:2018jhh})
\be
AIC_{cor}=AIC+\frac{2M(M+1)}{N-M-1}
\ee
For large samples as in our case ($N\gg M$) the correction term disappears but for small samples gives a more accurate answer.

The results for each candidate model are shown in Table \ref{comp} and the "preferred model" is the one which minimizes AIC and BIC. The absolute values of the AIC and BIC are not informative. Only the relative values between  different competing models are relevant. Hence when comparing one model versus the base/base-SH$0$ES we can use the model differences  $\Delta$AIC and $\Delta$BIC.

The differences $\Delta$AIC and $\Delta$BIC  with respect to the base/base-SH$0$ES model defined as
\be 
\Delta AIC=AIC_i-AIC_s=\Delta \chi_{min}^2+2 \Delta M
\label{daic}
\ee
\be 
\Delta BIC=BIC_i-BIC_s=\Delta \chi_{min}^2+\Delta M (ln{N})
\label{dbic}
\ee 
where the subindex i refers to value of AIC (BIC) for the model i and $AIC_s$ ($BIC_s$) is the value of AIC (BIC) for the base/base-SH$0$ES model. The resulting $\Delta$AIC and $\Delta$BIC are shown in Table \ref{comp} and in Fig. \ref{figbaric}. Note that a positive value of $\Delta$AIC or $\Delta$AIC means a preference for base/base-SH$0$ES model. 

According to the calibrated Jeffreys’ scales \cite{Jeffreys:1961} showed in the Table \ref{jefsc} (see also Refs. \cite{Liddle:2004nh,Nesseris:2012cq,BonillaRivera:2016use,Perez-Romero:2017njc,Camarena:2018nbr}) a range $0<|\Delta AIC|<2$ means that the two comparable  models have about the same support from the data, a range  $4<|\Delta AIC|<7$ means this support is considerably less for the model with the larger $AIC$ while for $|\Delta AIC|>10$ the model with the larger $AIC$ have no support i.e. the model is practically irrelevant. Similarly, for two competing models a range $0<|\Delta BIC|<2$ is regarded as weak evidence, a range $2<|\Delta BIC|<6$ is regarded as positive evidence, while for $|\Delta BIC|>6$ the evidence is strong against the model with the larger value. 

Ranking of the models considered according to AIC and BIC criteria are presented in Table \ref{rank}. Clearly, in terms of the AIC and BIC, fitting for two universal values of $R_W$ with global $M_H^W$ is the preferred model (case IV). Base/base-SH$0$ES model is considerable less supported by data with respect to the IV model ($\Delta$AIC) and  there is a positive/very strong evidence against it ($\Delta$BIC). We attribute the difference between $\Delta$AIC and $\Delta$BIC for the models considered to the fact that the BIC penalizes additional parameters more strongly than the AIC as inferred by the Eqs. (\ref{aic}) and (\ref{bic}) for the used dataset with $ln N>2$  (see Refs. \cite{Liddle:2004nh,Arevalo:2016epc,Rezaei:2019xwo}).

\section{Transition as a possible solution of Hubble tension}
\label{Transition as a possible solution of Hubble tension}

In this section, we investigate whether the existence of transition of the Cepheid parameters can impact on the inferred value of Hubble constant $H_0$ and its corresponding uncertainties. Using the best fit values of SnIa absolute magnitude $M_B$ the Hubble constant is given by
\be
H_0=10^{0.2 M_B+\alpha_B+5}
\label{ho1}
\ee
where the term $\alpha_B$ is the intercept of the SnIa magnitude-redshift relation defined as \cite{Riess:2016jrr}

\begin{align}
&\alpha_B =\log \left[cz\left(1+\frac{1}{2}(1-q_0)z \right.\right.\nonumber \\
&\left.\left.-\frac{1}{6}(1-q_0-3q_0^2+j_0)z^2+\mathcal{O}(z^3)\right) \right]-0.2m_B
\label{defabp}
\end{align}
where $q_0\equiv -\frac{1}{H_0^2}\frac{d^2a(t)}{dt^2}\Big|_{t=t_0}$ and $j_0\equiv \frac{1}{H_0^3}\frac{d^3a(t)}{dt^3}\Big|_{t=t_0}$ are the deceleration and jerk parameters respectively.

The intercept $\alpha_B$ using $217$ observed SnIa at redshifts $0.023 <z <0.15$ with the deceleration and jerk parameters set to $q_0= -0.55$ and $j_0= 1$ is determined to be $\alpha_B =0.71273\pm0.00176$ by Ref. \cite{Riess:2016jrr}.

Alternatively, using the the best fit value of degenerate combination ${\cal M}=23.803\pm 0.007$  as derived by Ref. \cite{Kazantzidis:2020tko} for full Pantheon dataset in Eq. (\ref{combM}) the Hubble constant can be estimated   
\be
H_0= c\, 10^{0.2 (M_B-\mathcal{M})+5}
\label{ho2}
\ee

In Table \ref{comp} we show the  best fit value of SnIa absolute magnitude $M_B$ and the corresponding Hubble constant $H_0$ as derived using the Eq. (\ref{ho1}) and the Eq. (\ref{ho2}) (the values in the parentheses) for all cases studied in this paper. Also the one dimensional relative probability density values of $M_B$ and $H_0$ as derived using the Eq. (\ref{ho1})  and the Eq. (\ref{ho2}) (the values in the parentheses of table \ref{comp}) compared to that from the Planck CMB measurement assuming flat $\Lambda$CDM are shown in Fig. \ref{figprmb} and Fig. \ref{figprh0} respectively. 

Clearly, for all cases which break the assumption of universality of the parameters $R_W$ and $M_H^W$ (i.e. I, II, III, IV, V, VI) the best fit value of SnIa absolute magnitude $M_B$  and the derived values of $H_0$ decrease and become consistent with the corresponding predicted CMB best fit values. For the preferred  model (case IV) we obtain $H_0=68.06\pm 1.80$ $km$ $ s^{-1} Mpc^{-1}$  with Planck tension $ <1\sigma$. Therefore the transition in the Cepheid calibrator parameters at $D_c \simeq 16\,Mpc$ can provide a resolution of the Hubble tension.

\section{CONCLUSION-DISCUSSION}
\label{CONCLUSION-DISCUSSION}

In the present study we have used  Cepheid SnIa calibrator data to investigate the effects of variation of the Cepheid calibration empirical parameters. We have shown that models where such a variation is allowed are favored on the basis of model selection criteria AIC and BIC. The models that are consistently favored by both AIC and BIC involve a transition in either the color-luminosity parameter $R_W$ or the Cepheid absolute magnitude $M_H^W$, at a distance in the range between $10$ and $20\,Mpc$. In the context of a homogeneous Universe where the cosmological principle is respected this would be a transition in time between about $25\,Myrs$ and $70\,Myrs$ ago. Models involving a transition in $R_W$ are slightly favored over models where there is a transition in $M_H^W$. Both classes of models, lead to values of $H_0$ that are consistent with the CMB inferred values thus eliminating the Hubble tension. 

Such a transition of Cepheid parameters could be induced by a fundamental physics transition. The magnitude of the transition is consistent with the magnitude required for the resolution of the Hubble tension in the context of a fundamental gravitational transition occurring by a sudden increase of the  strength of the gravitational interactions $G_{eff}$ by about $10\%$ \cite{Marra:2021fvf} at a redshift $z_t\lesssim 0.01$ ($\lesssim 150$ million years). Such a transition would  abruptly increase the SnIa absolute magnitude by $\Delta M_B \simeq 0.2$  \cite{Alestas:2020zol,Marra:2021fvf} (from $M_B =-19.401\pm 0.027\, mag$ for $z > z_t$  \cite{Camarena:2019rmj} to $M_B=-19.244\pm 0.037 \,mag$ for $z < z_t$ \cite{Camarena:2019moy,Camarena:2021jlr}). The distance range/timescale corresponding to this transition is consistent with a recent analysis indicating a similar transition in the context of the Tully-Fisher data \cite{Alestas:2021nmi}. 

An alternative origin of the observed effect is based on a scenario where the parameter $R_W$ could vary across different sightlines and different galaxy distances, morphologies, environments and properties. Dust extinction differences between galaxies could be the origin for a systematic ”mass step” (at $\sim 10^{10}\,M_{\odot}$) in the data \cite{Brout:2020msh,2021arXiv210506236J}: after standardization, SnIa in a high-mass galaxy  appear brighter than those in a low-mass galaxy  \cite{1995AJ....109....1H,SupernovaCosmologyProject:2002eym,Childress:2013xna,Scolnic:2017caz,DES:2020xhr,DES:2020mzh}. Such an alternative scenario is testable using the methods presented here and it could also lead to a resolution of the Hubble tension. Such an extension is beyond the goals of the present analysis.

Other interesting extensions of the present analysis include the following:
\begin{itemize}
\item
The search for transition in other parameters that can be constrained using the Cepheid data (for example the SnIa absolute magnitude $M_B$ and its effect on the estimation of $H_0$).
\item
If the transition is physical and due to fundamental gravitational physics it would be interesting to search for possible constraints or remnants of such a transition in the solar system. For example what could be the relevance of such a transition on the observed increased rate of impactors on Earth and Moon by a factor of 2-3 during the past 290 Myrs \cite{Mazrouei253} which may also be related with the Cretaceous–Paleogene (K-Pg) extinction event (also known as the Cretaceous–Tertiary (K–T) extinction)  (e.g. \cite{Schulte1214}) that  eliminated several species on Earth about 66 Myrs ago \cite{Renne684} including dinosaurs (e.g. \cite{Alroy11536}). According to the impact hypothesis or Alvarez hypothesis \cite{Alvarez1095,1983PNAS...80..627A} this extinction event was caused by an asteroid impact producing the $\sim 200\,km$ wide Chicxulub impact crater.  
\item
It would be interesting to search for a similar transition in the other SnIa calibrator such as the Tip (a sharp discontinuity) of the Red Giant Branch (TRGB) in the  Hertzsprung-Russell diagram  \cite{Beaton:2016nsw}. The Red Giant stars have nearly exhausted the hydrogen in their cores and have just began helium burning by the triple-a process (helium flash phase). The brightness of TRGB stars can be standardized using DEBs combined with parallax calibration. They can serve as excellent alternative standard candles \cite{Cassisi:1997fr} visible in the local Universe for the subsequent calibration of SnIa \cite{2015ApJ...807..133J,Jang:2017dxn,Hatt:2018opj,Hatt:2018zfv} and thereby provide an independent determination of the Hubble constant $H_0$ \cite{Freedman:2019jwv,Freedman:2020dne}.
\end{itemize}

The indicated transition at a distance of about $10-20\,Mpc$ could be interpreted as violating the cosmological principle according to which the distance of any galaxy from us should not impact its properties.  The cosmological principle however is not necessarily violated in the transition models because a spatial transition can not be observationally distinguished from a temporal transition. If the transition is temporal and occurred at a specific time then there is no violation of the cosmological principle.

Even if the transition is spatial it could be interpreted as a result of a first order phase transition occurring very recently due to a decay of the false vacuum \cite{Coleman:1977py}. Then we could live in $20\,Mpc$ true vacuum bubble where a first order scalar-tensor physics transition has occurred. If the bubble was created at recent cosmological times (e.g. last $100\,Myrs$) in the context of a decay of a false vacuum (see e.g. Ref. \cite{Coleman:1977py}) then we would not have been able to see the other true vacuum bubbles since light from them may not have reached us yet. Thus even in that case there would be no apparent large-scale violation of the cosmological principle. The phenomenology of such recent false vacuum decay in the context of scalar-tensor theories is another interesting extension of our analysis. In this context it may be shown that for a transition energy scale similar to the present Hubble constant the typical scale of the true vacuum bubbles produced would be $15-20\,Mpc$. 
Fine tuning questions also arise in the context of the indicated transition: 'What is special about the scale of $15-20\,Mpc$ where the transition signal appears to exist?'. In the context of a false vacuum decay bubble there is no more fine tuning than in the \lcdm. If we accept the scale of the cosmological constant and the fact that there is a first order phase transition to another vacuum of a similar energy scale ($\sim 0.002 \,eV$) then the predicted spatial scale of the produced bubbles is theoretically predicted to be about $15\,Mpc$. If we allow for some true vacuum bubble growth (they expand with the speed of light) it could increase to the scale of $20\,Mpc$. This generic result may be demonstrated as follows \cite{Patwardhan:2014iha,Coleman:1977py,Callan:1977pt,Doran:2006kp}: For a  very recent false vacuum decay with vacuum energy comparable to the cosmological constant the scale of the produced bubbles is 
\be 
R_b = \delta/H_0
\label{rbubble}
\ee
where $\delta$ depends logarithmically on the ratio of the Planck mass $M_P$ to the transition temperature energy scale $T_c=2.7^\circ K\simeq 2\times 10^{-4}eV$ as  \cite{Patwardhan:2014iha}
\be 
\delta \simeq \left[4 B_1 \ln\left(M_P/T_c\right) \right]^{-1}
\label{del}
\ee
where $B_1$ is a constant of $O(1)$. Using Eqs. (\ref{rbubble}) and (\ref{del}) with $H_0=70\,km\,s^{-1}\,Mpc^{-1}$ we obtain $R_b\simeq 15\,Mpc$ which is clearly within the range of transition scales favored by the Cepheid data by the present analysis and by the Tully-Fisher data as indicated by Ref. \cite{Callan:1977pt,Coleman:1977py,Doran:2006kp}. 

In conclusion the revolutionary improvement in the quality and quantity of data  from existing and upcoming missions/experiments raises the expectation of determining the origin of the existing transition effect shown in our analysis.  One possible origin would be the presence of systematic errors affecting the adopted calibration method. Alternatively, if the source of the demonstrated transition is physical it could lead to new cosmological physics beyond the standard model which may include a very recent false vacuum decay.\\

\section*{Numerical Analysis Files}

The numerical files for the reproduction of the figures can be found in the \href{https://github.com/FOTEINISKARA/Cepheid_SnIa_Calibrator_Data_Transition}{Cepheid SnIa Calibrator Data Transition} Github repository under the
MIT license.\\

\section*{Acknowledgements}
We thank Adam Riess and Dan Scolnic for interesting and useful comments. This research is co-financed by Greece and the European Union (European Social Fund-ESF) through the Operational Programme "Human Resources Development, Education and Lifelong Learning 2014-2020" in the context of the project MIS 5047648. \\

\appendix
\section{MATRICES OF SYSTEM OF EQUATIONS}
\label{SYSTEM OF EQUATIONS}

In this Appendix we present the schematic form of the error matrix $\bf{C}$, the matrix of measurements $\bf{Y}$,  the matrix of parameters $\bf{X}$ and  the equation (or design) matrix $\bf{A}$ used in the system of equations of our analysis (see Eqs. (\ref{syst}), (\ref{chi2}), (\ref{bfpar}) and (\ref{covmat}) in Section \ref{SEARCH FOR TRANSITION}).

\begin{widetext}
The schematic form of the error matrix $\bf{C}$ is
\be
\bf{C}=
\setcounter{MaxMatrixCols}{50}
\begin{pmatrix}
 \sigma^2_{MW,j} & 0 & \ldots  & & & & & & & & 0 \\
0 & \sigma^2_{tot,j}  &0 & \ldots & & & & & & & \vdots \\
0 & 0  &\sigma^2_{LMC,j} & 0 &\ldots & & & & &&  \vdots \\
\vdots & &  &\ddots & & & & & & &  \\
 & & & & \sigma^2 \left(\mu_{N4258} \right) & & & & & &  \\
 & & & & & \sigma^2 \left(\mu_{LMC} \right)& & & & &  \\
 & & & & & & \sigma^2 \left(m_{B,1} \right) & & & &  \\
 & & & & & & & \ddots  & & & \\
 0 & & & & & & & & & & \sigma^2 \left(m_{B,19} \right)
\end{pmatrix}
\ee
where $\sigma \left(\mu_{LMC} \right)= 0.0263$ is the error of the distance modulus to the LMC reported by Ref. \cite{2019Natur.567..200P}, $\sigma \left(\mu_{N4258} \right)= 0.032$ is the error of the distance modulus to the NGC 4258 reported by Ref. \cite{Reid:2019tiq}  and $\sigma \left(m_{B,k}\right)$  $(k=1,..,19)$ are the errors of SnIa B-band magnitudes obtained from Table 5 in Ref. \cite{Riess:2016jrr} (see our Table \ref{tab:hossn} of the Appendix \ref{DATA USED IN THE ANALYSIS}).

For MW $jth$ Cepheid we use a total statistical uncertainty arising from the quadrature sum of four terms (higher order terms $\mathcal{O}(zp/\pi_j)^2$ are negligible)
\be
\sigma_{MW,j}^2 =\sigma^2 \left( m_{H,j}\right)+\left(\frac{5}{\ln{10}}\frac{1}{\pi_j}\right)^2\sigma^2(\pi_j)+ R_W^2 \sigma^2(V_j)+R_W^2 \sigma^2(I_j)
\ee
and for LMC Cepheids we use a total statistical uncertainty arising from the quadrature sum of four terms:
\be
\sigma^2_{LMC,j} =\sigma^2 \left( m_{H,j}\right)+\sigma^2(\mu)+ R_W^2 \sigma^2(V_j)+R_W^2 \sigma^2(I_j)
\ee
We see that for the MW and LMC Cepheids where the color errors are provided by SH$0$ES team (see in Table 1 of Ref. \cite{Riess:2020fzl} and and in Table 2 of Ref. \cite{Riess:2019cxk} or our Table \ref{tab:mwcep} and our Table \ref{tab:lmccep} of the Appendix \ref{DATA USED IN THE ANALYSIS} respectively) we have included them in the error matrix $\bf{C}$ in the proper 2D manner (i.e. 2D fit including errors in both  $\bf{Y}$ and $\bf{X}$ "axes").

For the $jth$ Cepheid in the $ith$ galaxy (other than MW and LMC) where SH$0$ES does not provide separate color errors we use  a total statistical uncertainty  $\sigma_{tot}$ arising from the quadrature sum of four terms: NIR photometric error, color error, intrinsic width and random-phase as derived by SH$0$ES team and shown in column 8 of  Table 4 of Ref. \cite{Riess:2016jrr} 
\be
\sigma_{tot,j}^2=\sigma_{sky}^2+\sigma_{col}^2+\sigma_{int}^2+(f_{pf}\sigma_{ph})^2
\ee
These total statistical uncertainties  $\sigma_{tot}$ are shown in our Table \ref{tab:hoscep} of the Appendix \ref{DATA USED IN THE ANALYSIS}. Note that even though the color errors  are included implicitly in our  fit,  in order to make a full 2D fit we will need the separate color errors which are not publicly available for these Cepheids by SH$0$ES team. 

The schematic form of the matrix of measurements $\bf{Y}$ and  the matrix of parameters $\bf{X}$ are
\end{widetext}
\begin{widetext}

\centering
\be
\begin{tabular}{ccc}
\(\bf {Y}=
\begin{pmatrix}
m_{\pi,j}\\
m_{H,1,j}\\
\vdots\\
m_{H,19,j}\\
m_{H,N4258,j}\\
m_{H,M31,j}\\
m_{H,LMC,j}\\
\mu_{N4258}\\
\mu_{LMC}\\
m_{B,1}\\
\vdots\\
m_{B,19}
\end{pmatrix}\)  
 &,   &
\(
\bf{X}=
\begin{pmatrix}
R_{W,LMC}\\
R_{W,MW}\\
R_{W,1}\\
\vdots\\
R_{W,19}\\
R_{W,N4258}\\
R_{W,M31}\\
\mu_1\\
\vdots\\
\mu_{19}\\
\mu_{N4258}\\
\mu_{M31}\\
\mu_{LMC}\\
M_{H,MW}^W\\
M_{H,1}^W\\
\vdots\\
M_{H,19}^W\\
M_{H,N4258}^W\\
M_{H,M31}^W\\
M_{H,LMC}^W\\
b_W^s\\
b_W^l\\
Z_W\\
zp\\
M_B
\end{pmatrix}\)  \\
&&
\end{tabular}
\ee
\end{widetext}

\begin{widetext}
The schematic form of  the equation (or design) matrix $\bf{A}$ is
 \be
\begin{turn}{90}
\(\bf{A}=
\setcounter{MaxMatrixCols}{50}
\begin{pmatrix}
0&(V-I)_{MW,1}&0 &. & .& .& & & &0 &. &. &.&&&&1&0&.&.&.&&&[P]_{MW,1}^s&[P]_{MW,1}^l&[M/H]_1&\frac{-5\pi_1^{-1}}{\ln{10}}&0\\
\vdots\\
0&(V-I)_{MW,N}&0 &. & .& .& & & &0 &. &. &.&&&&1&0&.&.&.&&&[P]_{MW,N}^s&[P]_{MW,N}^l&[M/H]_{19}&\frac{-5\pi_{N}^{-1}}{\ln{10}}&0\\
0&0&(V-I)_{1,1}&0 &. & .& .& & &1&0&.&. &&&&0&1&0&.&.&&&[P]_{1,1}^s&[P]_{1,1}^l&[M/H]_{1,1}&0&0\\
\vdots\\
0&0&(V-I)_{1,N}&0 &. & .& .& & &1&0&.&. &&&&0&1&0&.&.&&&[P]_{1,N}^s&[P]_{1,N}^l&[M/H]_{1,N}&0&0\\
0&0&0&(V-I)_{2,1}&0 &. & .& .& & 0&1&0&.&. &&&0&0&1&0&.&&&[P]_{2,1}^s&[P]_{2,1}^l&[M/H]_{2,1}&0&0\\
\vdots\\
0&0&0&(V-I)_{2,N}&0 &. & .& .& &0 &1&0&.&. &&&0&0&1&0&.&&&[P]_{2,N}^s&[P]_{2,N}^l&[M/H]_{2,N}&0&0\\
\vdots\\
0&0&0&.&.&. & & &(V-I)_{M31,1} & 0&.&.& & &1&0&0&.&.&. &&1&0&[P]_{M31,1}^s&[P]_{M31,1}^l&[M/H]_{M31,1}&0&0\\
\vdots\\
0&0&0&.&.&. & & &(V-I)_{M31,N} & 0&.&.& & &1&0&0&.&.&.& &1&0&[P]_{M31,N}^s&[P]_{M31,N}^l&[M/H]_{M31,N}&0&0\\
(V-I)_{LMC,1}&0&0&.&.&. & & & & 0&.&.&& &0&1&0&.&.&.& &0&1&[P]_{LMC,1}^s&[P]_{LMC,1}^l&[M/H]_{LMC,1}&0&0\\
\vdots\\
(V-I)_{LMC,N}&0&0&.&.&. & & & & 0&.&.&&&0 &1&0&.&.&.& &0&1&[P]_{LMC,N}^s&[P]_{LMC,N}^l&[M/H]_{M31,N}&0&0\\
0&.& .& & && & & & 0&.&.&&1&0 &0&0&.&.&& &  & &  & & & \\
0&.& .& & && & & & 0&.&.&&0&0 &1&0&.&.&& &  & &  & & & \\
0&.& .& & && & & &1&0&.&.& &&&&&& &&  && &  & &&1 \\
0&.& .& & && & & &0&1&0&.&. &&&&&& &&  && &  & &&1 \\ 
\vdots& &  & & && & & &  & & & & &  & & & & && &  & &  & & &&\vdots\\
0&.& .& & && & & & 0&.&.&1&0&0 &.&.& & && &  & &  & & && 1
\end{pmatrix}\)
\end{turn}
\ee

\section{DATA USED IN THE ANALYSIS}
\label{DATA USED IN THE ANALYSIS}

In this appendix we present the data used in the
analysis.\\

\begin{longtable}{ | c | c | c | c | c | c | c |c|c|c|c|c|c|c|c| }
\caption{Photometric data for MW Cepheids from Table 1 in Ref. \cite{Riess:2020fzl}.} 
\label{tab:mwcep}\\
% header and footer information
\hline
 & &&&&&&&&&&&&&\\
Cepheid & log P & F 555W & $\sigma$ & F 814W & $\sigma$ & F 160W&$\sigma$&$m_H^W$ & $\sigma$   & $[Fe/H]$  & $\pi$& $\sigma$  & $\pi_{EDR3}$ &$\sigma$ \\ 
 & &[mag]&[mag] &[mag] &[mag]&[mag]&[mag]&[mag]&[mag]&&[mas]&[mas]&[mas]&[mas]\\
  & &&&&&&&&&&&&&\\
\hline
\hline
 & &&&&&&&&&&&&&\\
AA-GEM&1.053&9.9130&0.029&8.542&0.025&7.348&0.017&6.860&0.023&-0.080&0.259&0.008&0.311&0.019\\
AD-PUP&1.133&10.015&0.028&8.675&0.023&7.488&0.020&7.011&0.024&-0.060&0.214&0.006&0.254&0.018\\
AQ-CAR&0.990&8.9836&0.020&7.854&0.009&6.766&0.007&6.373&0.011&0.013&0.354&0.010&0.361&0.017\\
AQ-PUP&1.479&8.8671&0.018&7.120&0.014&5.487&0.013&4.859&0.016&0.060&0.340&0.010&0.294&0.025\\
BK-AUR&0.903&9.5609&0.036&8.220&0.038&7.015&0.021&6.539&0.029&0.070&0.371&0.011&0.426&0.016\\
BN-PUP&1.136&10.051&0.033&8.505&0.017&7.198&0.015&6.642&0.021&0.030&0.251&0.007&0.301&0.016\\
CD-CYG&1.232&9.1207&0.011&7.468&0.012&5.900&0.012&5.307&0.014&0.120&0.398&0.011&0.394&0.018\\
CP-CEP&1.252&10.757&0.015&8.638&0.052&6.871&0.022&6.095&0.030&0.050&0.270&0.008&0.279&0.022\\
CR-CAR&0.989&11.750&0.019&9.973&0.018&8.384&0.014&7.736&0.017&-0.080&0.190&0.005&0.194&0.016\\
CY-AUR&1.141&12.052&0.012&9.953&0.020&8.106&0.025&7.334&0.027&-0.150&0.183&0.006&$\ldots$ &$\ldots$\\
DD-CAS&0.992&10.036&0.007&8.523&0.011&7.108&0.012&6.566&0.013&0.160&0.319&0.009&0.346&0.014\\
DL-CAS&0.903&9.1059&0.019&7.569&0.022&6.238&0.018&5.689&0.021&0.050&0.550&0.016&$\ldots$ &$\ldots$\\
DR-VEL&1.049&9.7083&0.034&7.770&0.020&6.183&0.021&5.479&0.026&0.024&0.488&0.015&0.520&0.015\\
GQ-ORI&0.935&8.7199&0.020&7.632&0.024&6.523&0.032&6.146&0.034&0.250&0.418&0.013&0.408&0.023\\
HW-CAR&0.964&9.2782&0.016&8.007&0.013&6.798&0.005&6.350&0.009&0.060&0.370&0.010&0.397&0.013\\
KK-CEN&1.086&11.598&0.017&9.862&0.021&8.292&0.015&7.660&0.018&0.210&0.167&0.005&0.152&0.017\\
KN-CEN&1.532&10.062&0.023&7.924&0.017&5.856&0.006&5.076&0.013&0.550&0.273&0.008&0.251&0.020\\
RW-CAM&1.215&8.8673&0.015&7.044&0.014&5.451&0.021&4.794&0.022&0.080&0.519&0.015&$\ldots$ &$\ldots$\\
RW-CAS&1.170&9.3719&0.021&7.863&0.016&6.483&0.022&5.944&0.024&0.280&0.322&0.010&0.334&0.021\\
RY-CAS&1.084&10.075&0.019&8.333&0.040&6.715&0.010&6.085&0.020&0.320&0.342&0.010&0.359&0.016\\
RY-SCO&1.308&8.2067&0.012&6.206&0.010&4.408&0.010&3.685&0.012&0.010&0.757&0.021&0.764&0.035\\
RY-VEL&1.449&8.5234&0.036&6.757&0.016&5.211&0.017&4.576&0.023&0.090&0.403&0.012&0.376&0.023\\
S-NOR&0.989&6.5779&0.011&5.410&0.012&4.391&0.012&3.990&0.014&0.100&1.054&0.030&1.099&0.024\\
S-VUL&1.839&9.1668&0.008&6.862&0.012&4.885&0.010&4.043&0.011&0.090&0.287&0.008&0.237&0.022\\
SS-CMA&1.092&10.121&0.012&8.444&0.008&6.894&0.011&6.289&0.012&0.012&0.315&0.009&0.308&0.014\\
SV-PER&1.046&9.2186&0.016&7.760&0.014&6.435&0.027&5.916&0.028&0.030&0.400&0.012&$\ldots$ &$\ldots$\\
SV-VEL&1.149&8.7316&0.026&7.302&0.009&6.024&0.010&5.517&0.015&0.090&0.411&0.012&0.434&0.019\\
SV-VUL&1.653&7.2675&0.047&5.648&0.033&4.214&0.027&3.639&0.035&0.110&0.457&0.015&0.402&0.023\\
SY-NOR&1.102&9.8284&0.023&7.925&0.038&6.214&0.013&5.523&0.022&0.230&0.435&0.013&$\ldots$ &$\ldots$\\
SZ-CYG&1.179&9.6209&0.013&7.756&0.017&6.004&0.008&5.329&0.012&0.150&0.426&0.012&0.445&0.014\\
T-MON&1.432&6.0680&0.023&4.828&0.016&3.725&0.021&3.298&0.024&0.040&0.749&0.022&0.745&0.057\\
U-CAR&1.589&6.3852&0.038&4.967&0.023&3.768&0.019&3.272&0.026&0.250&0.589&0.018&0.561&0.025\\
UU-MUS&1.066&9.9212&0.024&8.457&0.025&7.108&0.010&6.584&0.017&0.190&0.282&0.008&0.306&0.013\\
V-339-CEN&0.976&8.8402&0.024&7.321&0.016&5.990&0.024&5.448&0.026&-0.080&0.557&0.017&0.568&0.023\\
V-340-ARA&1.318&10.460&0.024&8.554&0.014&6.808&0.012&6.115&0.016&-0.080&0.245&0.007&0.239&0.022\\
VW-CEN&1.177&10.379&0.031&8.718&0.023&7.158&0.010&6.558&0.018&0.410&0.238&0.007&0.260&0.017\\
VX-PER&1.037&9.4589&0.008&7.906&0.006&6.470&0.009&5.914&0.010&0.030&0.407&0.011&0.392&0.019\\
VY-CAR&1.276&7.6162&0.014&6.253&0.007&4.991&0.004&4.513&0.007&0.080&0.539&0.015&0.565&0.018\\
VZ-PUP&1.365&9.7715&0.033&8.262&0.022&6.931&0.017&6.390&0.023&-0.010&0.200&0.006&0.220&0.016\\
WX-PUP&0.951&9.1909&0.030&7.944&0.012&6.807&0.010&6.368&0.016&-0.010&0.376&0.011&0.387&0.017\\
WZ-SGR&1.339&8.2021&0.012&6.481&0.013&4.858&0.009&4.242&0.011&0.280&0.547&0.015&0.612&0.031\\
X-CYG&1.214&6.5295&0.020&5.230&0.049&4.080&0.033&3.629&0.039&0.160&0.883&0.029&0.910&0.022\\
X-PUP&1.414&8.6949&0.019&7.128&0.010&5.628&0.008&5.069&0.012&0.020&0.341&0.010&0.397&0.022\\
XX-CAR&1.196&9.4627&0.027&8.067&0.015&6.833&0.022&6.337&0.025&0.010&0.264&0.008&0.305&0.016\\
XY-CAR&1.095&9.4660&0.011&7.927&0.009&6.455&0.006&5.904&0.008&0.012&0.375&0.010&0.390&0.015\\
XZ-CAR&1.221&8.7725&0.017&7.217&0.006&5.770&0.007&5.215&0.010&0.026&0.425&0.012&0.473&0.020\\
YZ-CAR&1.259&8.8644&0.016&7.401&0.007&5.991&0.013&5.471&0.015&-0.030&0.359&0.010&0.358&0.020\\
YZ-SGR&0.980&7.4662&0.021&6.176&0.014&5.103&0.020&4.653&0.022&0.120&0.786&0.023&0.860&0.027\\
Z-LAC&1.037&8.5686&0.022&7.157&0.015&5.917&0.018&5.417&0.021&0.070&0.509&0.015&0.510&0.023\\
AG-CRU&0.584&8.3175&0.013&7.307&0.011&6.414&0.027&6.068&0.028&0.020&0.748&0.023&0.758&0.022\\
AP-PUP&0.706&7.4560&0.016&6.412&0.014&5.534&0.027&5.177&0.028&-0.020&0.941&0.029&0.924&0.022\\
AP-SGR&0.704&7.1056&0.028&6.036&0.013&5.094&0.027&4.729&0.030&0.160&1.145&0.035&1.217&0.026\\
BF-OPH&0.609&7.5091&0.018&6.347&0.010&5.374&0.027&4.972&0.028&0.110&1.184&0.036&1.189&0.026\\
BG-VEL&0.840&7.7827&0.010&6.299&0.009&5.054&0.019&4.529&0.020&0.040&1.033&0.030&1.045&0.019\\
ER-CAR&0.888&6.9095&0.011&5.916&0.012&5.078&0.027&4.742&0.028&0.120&0.867&0.026&0.869&0.016\\
R-CRU&0.765&6.8479&0.017&5.856&0.016&4.984&0.027&4.649&0.028&0.100&1.088&0.033&1.078&0.031\\
R-MUS&0.876&6.4568&0.009&5.447&0.008&4.609&0.019&4.268&0.020&-0.110&1.117&0.033&1.076&0.019\\
R-TRA&0.530&6.7236&0.013&5.794&0.014&5.025&0.019&4.714&0.020&0.160&1.497&0.044&1.560&0.018\\
RV-SCO&0.783&7.1616&0.010&5.871&0.007&4.773&0.019&4.323&0.020&0.080&1.234&0.036&1.257&0.023\\
RX-CAM&0.898&7.8310&0.016&6.215&0.013&4.791&0.028&4.216&0.029&0.080&1.090&0.034&$\ldots$ &$\ldots$\\
RY-CMA&0.670&8.2358&0.015&7.111&0.013&6.045&0.027&5.656&0.028&0.140&0.787&0.024&0.825&0.032\\
S-CRUe&0.671&6.6700&0.050&5.698&0.011&4.843&0.027&4.516&0.033&0.080&1.335&0.042&1.342&0.026\\
S-TRA&0.801&6.5171&0.013&5.553&0.012&4.752&0.027&4.429&0.028&0.010&1.150&0.035&1.120&0.024\\
SS-SCT&0.565&8.3122&0.010&7.073&0.005&6.034&0.019&5.600&0.019&0.110&0.948&0.028&0.934&0.025\\
T-VEL&0.667&8.1205&0.009&6.915&0.007&5.839&0.019&5.419&0.020&-0.160&0.904&0.026&0.940&0.018\\
TX-CYG&1.168&9.6108&0.024&7.083&0.015&4.789&0.027&3.862&0.029&0.260&0.844&0.026&0.829&0.020\\
U-AQL&0.847&6.5396&0.019&5.168&0.029&4.115&0.027&3.636&0.030&0.140&1.531&0.047&$\ldots$ &$\ldots$\\
U-SGR&0.829&6.8864&0.018&5.388&0.011&4.143&0.027&3.615&0.028&0.140&1.588&0.049&1.605&0.025\\
V-CAR&0.826&7.4753&0.009&6.403&0.008&5.463&0.019&5.096&0.020&0.080&0.810&0.024&0.797&0.015\\
V-VEL&0.641&7.5198&0.013&6.555&0.010&5.693&0.027&5.366&0.028&0.000&0.951&0.029&0.953&0.019\\
V0386-CYG&0.721&9.8126&0.015&7.748&0.014&5.944&0.027&5.192&0.028&0.170&0.901&0.028&0.894&0.014\\
V0482-SCO&0.656&8.0697&0.013&6.773&0.013&5.697&0.027&5.242&0.028&0.019&0.982&0.030&0.993&0.028\\
V0636-SCO&0.832&6.8167&0.009&5.618&0.008&4.568&0.020&4.154&0.021&0.070&1.239&0.036&1.180&0.037\\
W-GEM&0.898&7.0841&0.057&5.899&0.018&4.863&0.027&4.454&0.036&-0.010&0.984&0.032&1.006&0.031	\\

\hline
\end{longtable}

\begin{longtable}{ | c | c | c  m{0.0001\textwidth} | c | c | c| c |c|c|c|c|c|cc| }
\caption{Photometric data for LMC Cepheids from Table 2 in Ref. \cite{Riess:2019cxk}.} 
\label{tab:lmccep}\\
% header and footer information
\hline
 & &&&&&&&&&&&&&\\
Cepheid & RA  & DEC && Geo &  log Period & F 555W&$\sigma$ & F 814W&$\sigma$&F 160W&$\sigma$ &$m_H^W$ & $\sigma$ &   \\ 
 & & & & & &[mag] &[mag] & [mag]&[mag] &[mag]  & [mag] & [mag]&[mag]& \\
 & &&&&&&&&&&&&&\\ 
\hline 
\hline
 & &&&&&&&&&&&&&\\
OGL0434&74.114583&-69.379611&&0.028&1.482&13.131&0.028&12.208&0.011&11.321&0.018&10.966&0.021&\\
OGL0501&74.462625&-69.958250&&0.034&1.367&13.623&0.022&12.693&0.012&11.770&0.021&11.406&0.023&\\
OGL0510&74.523208&-69.454333&&0.027&1.566&13.457&0.037&12.299&0.021&11.232&0.042&10.787&0.045&\\
OGL0512&74.545000&-69.949694&&0.033&1.595&13.134&0.025&12.005&0.017&11.038&0.017&10.598&0.020&\\
OGL0528&74.636583&-70.346028&&0.038&1.553&13.175&0.052&12.156&0.021&11.226&0.020&10.824&0.029&\\
OGL0545&74.696292&-70.061583&&0.034&1.199&14.414&0.045&13.349&0.010&12.311&0.018&11.895&0.025&\\
OGL0590&74.921417&-69.456111&&0.025&1.502&13.470&0.025&12.382&0.014&11.311&0.038&10.895&0.039&\\
OGL0594&74.937833&-69.493194&&0.025&0.828&15.279&0.012&14.352&0.011&13.525&0.030&13.171&0.030&\\
OGL0648&75.201500&-69.531861&&0.025&1.134&14.740&0.012&13.675&0.009&12.714&0.028&12.308&0.029&\\
OGL0683&75.353917&-70.071750&&0.031&1.166&14.446&0.009&13.350&0.009&12.481&0.020&12.056&0.021&\\
OGL0712&75.477375&-68.904028&&0.016&1.316&13.771&0.033&12.862&0.009&11.889&0.038&11.552&0.040&\\
OGL0716&75.503958&-68.922833&&0.016&1.085&14.958&0.012&13.895&0.011&12.938&0.023&12.541&0.024&\\
OGL0727&75.542667&-69.539917&&0.023&1.161&14.208&0.010&13.231&0.010&12.375&0.033&12.004&0.034&\\
OGL0757&75.629333&-69.397056&&0.021&0.924&15.153&0.012&14.214&0.012&13.364&0.023&13.010&0.024&\\
OGL0770&75.714750&-68.784806&&0.013&1.035&14.698&0.012&13.757&0.010&12.913&0.029&12.566&0.030&\\
OGL0787&75.787750&-69.223333&&0.018&1.243&14.243&0.009&13.186&0.009&12.281&0.042&11.884&0.042&\\
OGL0798&75.848542&-69.000889&&0.015&1.029&14.891&0.011&13.860&0.009&12.934&0.020&12.550&0.021&\\
OGL0800&75.854417&-68.772500&&0.012&1.101&14.541&0.018&13.584&0.011&12.712&0.043&12.359&0.044&\\
OGL0812&75.908167&-69.063083&&0.015&1.086&14.670&0.011&13.691&0.010&12.733&0.024&12.369&0.025&\\
OGL0819&75.942333&-68.876778&&0.013&1.348&14.162&0.014&13.013&0.010&11.987&0.029&11.560&0.030&\\
OGL0821&75.956250&-68.934083&&0.014&1.411&13.770&0.009&12.714&0.010&11.706&0.024&11.314&0.025&\\
OGL0831&75.988583&-68.840056&&0.012&0.987&14.761&0.012&13.827&0.010&12.969&0.023&12.625&0.024&\\
OGL0844&76.064458&-69.026778&&0.014&1.235&13.967&0.019&13.010&0.011&12.146&0.034&11.791&0.035&\\
OGL0847&76.081833&-68.930306&&0.013&1.314&14.437&0.009&13.317&0.011&12.320&0.030&11.904&0.031&\\
OGL0848&76.087833&-68.728556&&0.010&1.203&14.262&0.013&13.341&0.011&12.444&0.027&12.107&0.028&\\
OGL0888&76.316875&-68.723472&&0.009&0.971&14.889&0.010&13.960&0.010&13.135&0.022&12.796&0.022&\\
OGL0915&76.424875&-68.851472&&0.010&0.868&15.125&0.013&14.216&0.011&13.359&0.021&13.027&0.022&\\
OGL0936&76.504708&-68.627361&&0.007&0.920&15.046&0.015&14.101&0.012&13.260&0.023&12.917&0.024&\\
OGL0949&76.570375&-68.676028&&0.007&1.111&14.519&0.016&13.549&0.011&12.670&0.018&12.317&0.019&\\
OGL0966&76.699875&-70.037056&&0.024&1.676&12.944&0.026&11.861&0.014&10.896&0.023&10.483&0.025&\\
OGL0969&76.720250&-68.723639&&0.007&1.104&14.727&0.013&13.691&0.012&12.725&0.028&12.347&0.029&\\
OGL0970&76.720708&-68.659889&&0.006&1.242&14.369&0.014&13.249&0.011&12.260&0.031&11.850&0.032&\\
OGL0975&76.742833&-68.611417&&0.006&1.101&14.628&0.015&13.617&0.011&12.749&0.023&12.382&0.024&\\
OGL0978&76.748792&-68.723972&&0.007&1.022&14.855&0.011&13.817&0.010&12.908&0.021&12.529&0.022&\\
OGL0986&76.782542&-68.888750&&0.009&1.492&13.471&0.018&12.403&0.008&11.445&0.045&11.053&0.045&\\
OGL0992&76.816583&-68.883500&&0.009&1.723&12.305&0.016&11.297&0.011&10.436&0.073&10.067&0.073&\\
OGL1001&76.840375&-68.338417&&0.002&1.160&14.464&0.009&13.447&0.009&12.485&0.023&12.119&0.024&\\
OGL1031&76.925542&-69.246694&&0.013&1.266&14.455&0.011&13.348&0.009&12.284&0.021&11.873&0.022&\\
OGL1058&77.076125&-68.779750&&0.006&1.482&13.564&0.016&12.452&0.008&11.467&0.021&11.060&0.022&\\
OGL1080&77.183292&-68.757778&&0.005&1.270&14.135&0.014&13.061&0.011&12.097&0.025&11.706&0.026&\\
OGL1109&77.316417&-68.741556&&0.005&1.074&14.520&0.008&13.592&0.009&12.744&0.053&12.410&0.053&\\
OGL1112&77.326458&-68.299556&&-0.000&0.899&14.713&0.010&13.907&0.011&13.134&0.053&12.853&0.054&\\
OGL1313&78.580750&-69.490056&&0.008&0.834&15.128&0.012&14.278&0.012&13.529&0.026&13.222&0.027&\\
OGL1374&78.857208&-69.340917&&0.005&0.838&15.386&0.011&14.438&0.012&13.582&0.031&13.240&0.031&\\
OGL1389&78.909875&-69.255500&&0.004&0.862&14.965&0.011&14.139&0.012&13.382&0.019&13.088&0.020&\\
OGL1411&78.991833&-69.712083&&0.009&0.897&15.324&0.010&14.324&0.011&13.468&0.026&13.102&0.026&\\
OGL1417&79.000917&-69.538167&&0.007&0.938&14.879&0.009&13.954&0.010&13.138&0.031&12.803&0.031&\\
OGL1424&79.016000&-69.247889&&0.003&0.830&15.663&0.013&14.622&0.015&13.686&0.030&13.310&0.031&\\
OGL1431&79.041083&-69.544306&&0.007&1.010&14.829&0.010&13.841&0.009&13.028&0.027&12.669&0.027&\\
OGL1463&79.228708&-69.330667&&0.003&0.876&15.098&0.011&14.171&0.011&13.339&0.024&13.007&0.025&\\
OGL1466&79.244417&-69.393889&&0.004&0.789&15.581&0.012&14.607&0.013&13.719&0.031&13.368&0.032&\\
OGL1490&79.353083&-69.349333&&0.003&0.912&14.690&0.014&13.876&0.010&13.160&0.023&12.872&0.024&\\
OGL1526&79.515792&-69.426639&&0.003&0.828&15.147&0.012&14.287&0.011&13.475&0.025&13.169&0.026&\\
OGL1539&79.592125&-69.363139&&0.002&1.130&14.636&0.016&13.620&0.011&12.671&0.019&12.306&0.021&\\
OGL1578&79.811667&-69.605028&&0.005&1.123&14.329&0.017&13.379&0.010&12.590&0.023&12.248&0.024&\\
OGL1587&79.865750&-69.508389&&0.003&1.334&14.151&0.008&12.959&0.009&11.925&0.034&11.491&0.034&\\
OGL1616&79.999708&-69.173722&&-0.001&1.191&14.894&0.017&13.693&0.012&12.631&0.033&12.198&0.034&\\
OGL1637&80.095833&-69.038194&&-0.003&1.504&13.337&0.025&12.251&0.014&11.314&0.035&10.928&0.036&\\
OGL1641&80.119292&-69.025500&&-0.003&1.144&14.180&0.010&13.272&0.010&12.428&0.026&12.111&0.027&\\
OGL1647&80.155792&-69.515722&&0.002&0.939&14.833&0.019&13.957&0.015&13.157&0.028&12.846&0.030&\\
OGL1677&80.301958&-69.052111&&-0.004&1.372&13.717&0.013&12.716&0.017&11.816&0.020&11.463&0.022&\\
OGL1862&81.056042&-69.500444&&-0.001&1.118&14.868&0.013&13.763&0.010&12.753&0.028&12.357&0.028&\\
OGL1939&81.370042&-69.912361&&0.002&0.782&15.498&0.021&14.583&0.015&13.683&0.029&13.356&0.031&\\
OGL1940&81.370625&-69.834194&&0.001&0.972&16.659&0.022&15.055&0.017&13.509&0.026&12.917&0.028&\\
OGL1941&81.372000&-69.920167&&0.003&0.832&15.642&0.015&14.538&0.013&13.593&0.029&13.193&0.030&\\
OGL1945&81.381000&-69.834361&&0.001&0.885&16.166&0.020&14.832&0.016&13.610&0.023&13.123&0.025&\\
OGL1994&81.594042&-69.602056&&-0.002&0.889&14.779&0.022&13.946&0.011&13.229&0.019&12.939&0.022&\\
OGL2012&81.708292&-69.764667&&-0.000&0.872&15.061&0.011&14.136&0.010&13.327&0.025&13.000&0.025&\\
OGL2019&81.732917&-69.980222&&0.002&1.448&13.615&0.018&12.581&0.014&11.697&0.025&11.325&0.027&\\
OGL2043&81.845667&-69.849444&&0.000&0.867&15.246&0.011&14.320&0.012&13.495&0.038&13.167&0.039&\\

\hline
\end{longtable}

\begin{longtable}{ | c | c | c | c | c | c | c |c|cm{0.0001\textwidth}| }
\caption{WFC3-IR data for 1486 Cepheids in the anchor galaxy NGC 4258 and in the host galaxies from Table 4 in Ref. \cite{Riess:2016jrr}. An electronic version of the complete  table is available at \cite{githubfs}.} 
\label{tab:hoscep}\\
% header and footer information
\hline
&&&&&&&&&\\
Galaxy Name& $\alpha$ & $\delta$ & ID & P & $V-I$ & H &$\sigma_{tot}$&Z\footnote{$Z=12+\log(O/H)$}&  \\ 
  &   &   & [mag] & [days] &[mag]   & [mag] &[mag] &  [dex]&  \\ 
  &&&&&&&&&\\
\hline
\hline
&&&&&&&&&\\
N1309&50.513050&-15.412250&154632&38.10&1.08&25.46&0.22&8.582&							\\
N1309&50.514080&-15.405860&149317&39.31&1.34&25.31&0.42&8.722&							\\
N1309&50.537010&-15.412090&42756&39.42&1.14&25.28&0.24&8.793&							\\
N1309&50.536140&-15.385790&40303&39.50&1.00&25.19&0.24&8.758&							\\
N1309&50.534050&-15.388290&50270&39.83&1.22&25.56&0.26&8.866&							\\
N1309&50.520380&-15.397210&119907&40.84&1.40&24.57&0.55&8.974&							\\
N1309&50.516100&-15.386090&136479&41.02&1.15&25.07&0.18&8.646&							\\
N1309&50.531470&-15.406890&67093&41.77&1.23&24.45&0.45&9.037&							\\
N1309&50.528230&-15.408650&82654&41.94&1.05&24.24&0.50&8.987&							\\
N1309&50.540170&-15.394110&27150&42.11&1.03&24.49&0.28&8.860&							\\
N1309&50.533920&-15.386930&50545&44.14&1.11&25.34&0.57&8.825&							\\
N1309&50.534590&-15.392310&48826&44.32&1.65&24.54&0.34&8.975&							\\
N1309&50.522460&-15.384870&103930&44.60&0.89&24.95&0.22&8.739&							\\
N1309&50.520140&-15.408020&2099043&44.30&1.39&24.12&0.45&8.866&							\\
N1309&50.535980&-15.411540&47351&45.45&1.02&24.64&0.24&8.828&							\\
N1309&50.516850&-15.403490&2108877&47.59&1.34&24.78&0.53&8.841&							\\
N1309&50.513220&-15.403900&152242&47.82&0.95&24.76&0.33&8.715&							\\
N1309&50.514140&-15.404030&2117990&48.18&1.23&24.88&0.43&8.745&							\\
N1309&50.517600&-15.403870&134187&47.87&1.31&23.84&0.62&8.862&							\\
N1309&50.513500&-15.398810&150120&48.45&1.52&24.85&0.30&8.743&							\\
N1309&50.541640&-15.396450&22773&49.57&1.35&25.16&0.23&8.840&							\\
N1309&50.523070&-15.400960&105535&50.82&1.25&24.92&0.49&9.074&							\\
N1309&50.530380&-15.386750&66971&51.38&1.14&24.75&0.24&8.849&							\\
N1309&50.519300&-15.403510&126515&51.43&0.80&24.60&0.60&8.922&							\\
N1309&50.532360&-15.415870&65208&52.02&1.41&24.34&0.18&8.734&							\\
N1309&50.526790&-15.409480&89446&52.50&1.11&24.57&0.40&8.950&							\\
N1309&50.523400&-15.407320&105633&54.07&0.89&24.65&0.65&8.963&							\\
N1309&50.531490&-15.389480&62523&56.82&1.13&24.77&0.25&8.934&							\\
N1309&50.530220&-15.390540&68651&58.92&1.31&25.03&0.33&8.978&							\\
N1309&50.520230&-15.406500&122991&58.97&1.66&24.83&0.43&8.902&							\\
N1309&50.515030&-15.407650&145875&64.63&1.54&24.58&0.33&8.726&							\\
N1309&50.520350&-15.400180&120871&64.84&1.32&23.75&0.54&8.983&							\\
N1309&50.528410&-15.417520&83989&64.20&1.32&24.32&0.19&8.688&							\\
N1309&50.540600&-15.394620&25808&67.70&1.18&24.61&0.19&8.854&							\\
N1309&50.528080&-15.409230&83493&57.92&1.40&24.53&0.35&8.967&							\\
N1309&50.536050&-15.412330&47225&69.33&1.11&24.57&0.22&8.804&							\\
N1309&50.537880&-15.406430&37762&69.30&1.09&24.31&0.21&8.914&							\\
N1309&50.535540&-15.414100&49918&71.48&1.08&24.12&0.23&8.758&							\\
N1309&50.527070&-15.408140&87828&73.62&0.97&24.31&0.30&8.996&							\\
N1309&50.519710&-15.404750&124934&75.76&1.50&24.21&0.40&8.918&							\\
N1309&50.518570&-15.394630&127649&84.54&1.07&24.27&0.50&8.885&							\\
N1309&50.526840&-15.407730&88762&84.89&1.27&23.90&0.40&9.007&							\\
N1309&50.540840&-15.390800&24251&90.59&1.31&24.01&0.18&8.781&							\\
N1309&50.526100&-15.405700&91743&90.91&1.06&24.08&0.36&9.061&							\\
N1365&53.428340&-36.168310&111818&15.90&0.71&25.07&0.63&8.814&							\\
N1365&53.468230&-36.154760&140975&16.45&1.27&25.25&0.32&8.420&							\\
N1365&53.435410&-36.169560&123989&16.72&0.88&24.53&0.40&8.715&							\\
N1365&53.448980&-36.163290&132389&17.00&1.07&25.29&0.33&8.613&							\\
N1365&53.444760&-36.149450&103384&17.01&0.91&25.82&0.44&8.796&							\\
N1365&53.440820&-36.157210&111940&19.69&0.67&24.86&0.40&8.784&							\\
N1365&53.465100&-36.152740&136735&25.70&0.89&24.48&0.18&8.480&							\\
N1365&53.426190&-36.165250&101154&26.08&0.99&23.98&0.70&8.878&							\\
N1365&53.445370&-36.136230&63449&26.96&1.36&24.16&0.36&8.858&							\\
N1365&53.462490&-36.157290&138773&26.98&1.01&24.40&0.19&8.481&							\\
N1365&53.443040&-36.160620&120972&27.45&1.04&24.41&0.24&8.720&							\\
N1365&53.446770&-36.135500&65336&27.94&0.87&24.82&0.32&8.839&							\\
N1365&53.438940&-36.166810&124631&29.33&1.02&24.11&0.24&8.705&							\\
N1365&53.458140&-36.153810&130859&29.37&0.98&24.48&0.23&8.571&							\\
N1365&53.460390&-36.153990&133465&29.45&1.35&24.29&0.23&8.538&							\\
N1365&53.431590&-36.162200&105797&30.34&1.29&24.07&0.48&8.851&							\\
N1365&53.427620&-36.166050&106470&30.39&0.91&23.83&0.42&8.851&							\\
N1365&53.438980&-36.153370&100027&31.37&0.67&24.33&0.30&8.845&							\\
N1365&53.431200&-36.158650&94995&32.42&1.02&24.35&0.45&8.897&							\\
N1365&53.449180&-36.155960&122163&34.10&1.21&24.19&0.19&8.680&							\\
N1365&53.449310&-36.140060&87703&35.11&1.21&23.54&0.26&8.786&							\\
N1365&53.427540&-36.151730&61628&37.01&1.61&24.27&0.38&9.018&							\\
N1365&53.433170&-36.155410&90510&39.06&1.06&24.04&0.41&8.905&							\\
N1365&53.437550&-36.170950&128912&40.73&1.14&23.47&0.20&8.673&							\\
N1365&53.440360&-36.153510&103704&40.85&1.43&24.26&0.23&8.825&							\\
N1365&53.432210&-36.161310&104907&42.90&1.10&23.44&0.36&8.854&							\\
N1365&53.431420&-36.172780&123489&43.00&0.97&23.96&0.32&8.720&							\\
N1365&53.437390&-36.155420&101731&47.24&0.91&23.03&0.22&8.848&							\\
N1365&53.435220&-36.154740&94055&51.45&1.33&23.91&0.25&8.884&							\\
N1365&53.438630&-36.170270&129336&57.13&0.95&23.42&0.16&8.668&							\\
N1365&53.432110&-36.155840&88821&63.17&1.32&23.35&0.26&8.915&							\\
N1365&53.427080&-36.156250&75575&68.46&1.25&22.90&0.32&8.976&\\							
\hline
\end{longtable}

\begin{longtable}{ | c | c | c | c | c | c | c |c| }
\caption{Approximations for distance parameters from Table 5 in Ref. \cite{Riess:2016jrr}.} 
\label{tab:hossn}\\
% header and footer information
\hline
&&&&&&&\\
Host Galaxy & SnIa & $m_{B,i}^0+5\alpha_B$ & $\sigma$ & $\mu_{ceph}$ & $\sigma$ &  $M_{B,i}^0$&$\sigma$ \\ 
  &   & [mag]  & [mag] &[mag]  &[mag]   &[mag]  &[mag]  \\ 
 &&&&&&&\\ 
\hline
\hline
&&&&&&&\\
M101&  2011fe&  13.310&  0.117&  29.135&  0.045&  -19.389&  0.125\\
N1015&  2009ig&  17.015&  0.123&  32.497&  0.081&  -19.047&  0.147\\
N1309&  2002fk&  16.756&  0.116&  32.523&  0.055&  -19.331&  0.128\\
N1365&  2012fr&  15.482&  0.125&  31.307&  0.057&  -19.390&  0.137\\
N1448&  2001el&  15.765&  0.116&  31.311&  0.045&  -19.111&  0.125\\
N2442&  2015F&  15.840&  0.142&  31.511&  0.053&  -19.236&  0.152\\
N3021&  1995al&  16.527&  0.117&  32.498&  0.090&  -19.535&  0.147\\
N3370&  1994ae&  16.476&  0.115&  32.072&  0.049&  -19.161&  0.125\\
N3447&  2012ht&  16.265&  0.124&  31.908&  0.043&  -19.207&  0.131\\
N3972&  2011by&  16.048&  0.116&  31.587&  0.070&  -19.103&  0.136\\
N3982&  1998aq&  15.795&  0.115&  31.737&  0.069&  -19.507&  0.134\\
N4038&  2007sr&  15.797&  0.114&  31.290&  0.112&  -19.058&  0.160\\
N4424&  2012cg&  15.110&  0.109&  31.080&  0.292&  -19.534&  0.311\\
N4536&  1981B&  15.177&  0.124&  30.906&  0.053&  -19.293&  0.135\\
N4639&  1990N&  15.983&  0.115&  31.532&  0.071&  -19.113&  0.135\\
N5584&  2007af&  16.265&  0.115&  31.786&  0.046&  -19.085&  0.124\\
N5917&  2005cf&  16.572&  0.115&  32.263&  0.102&  -19.255&  0.154\\
N7250&  2013dy&  15.867&  0.115&  31.499&  0.078&  -19.196&  0.139\\
U9391&  2003du&  17.034&  0.114&  32.919&  0.063&  -19.449&  0.130\\

\hline
\end{longtable}

\end{widetext}

\raggedleft
\bibliography{bibliography}

\end{document}